\RequirePackage{booktabs} 

\documentclass[pdflatex,sn-chicago,10pt]{sn-jnl}

\usepackage{graphicx}%
\usepackage{multirow}%
\usepackage{amsmath,amssymb,amsfonts}%
\usepackage{amsthm}%
\usepackage[title]{appendix}%
\usepackage{xcolor}%
\usepackage{textcomp}%
\usepackage{manyfoot}%
\usepackage{booktabs}%
\usepackage{algorithm}%
\usepackage{algorithmicx}%
\usepackage{algpseudocode}%
\usepackage{listings}%
\usepackage{multirow}
\usepackage{float}

\usepackage[normalem]{ulem} 

\usepackage{enumitem}

\theoremstyle{thmstyleone}%
\theoremstyle{thmstyletwo}%

\theoremstyle{thmstylethree}%

\begin{document}

\setcounter{page}{1}








\author[1]{\fnm{Stefano Maria} \sur{Iacus}}\email{siacus@iq.harvard.edu}

\author*[2]{\fnm{Giuseppe} \sur{Porro}}\email{giuseppe.porro@uninsubria.it}

\affil*[1]{\orgdiv{Institute for Quantitative Social Science}, \orgname{Harvard University}, \orgaddress{\street{1737 Cambridge Street}, \city{Cambridge}, \postcode{100190}, \state{Massachusetts}, \country{USA}}}

\affil[2]{\orgdiv{Department of Law, Economics and Culture}, \orgname{University of Insubria}, \orgaddress{\street{Via S. Abbondio, 12}, \city{Como}, \postcode{22100}, \country{Italy}}}
 
\title{Two Americas of Well-Being: Divergent Rural–Urban Patterns of Life Satisfaction and Happiness from 2.6 B Social Media Posts}

\abstract{
Using 2.6 billion geolocated {tweets} (2014--2022) and a 
fine-tuned generative language model, we construct county-level indicators 
of \emph{life satisfaction} and \emph{happiness} for the United States. 
{We document an \emph{apparent rural--urban paradox}: even 
in unadjusted county-level means, rural counties express higher 
\emph{life satisfaction} while urban counties exhibit greater 
\emph{happiness}. This opposite gradient persists and is further 
characterized} once the two are treated as distinct layers 
of subjective well-being, evaluative vs.\ hedonic, showing that each maps 
differently onto place, politics, and time. {Democratic-leaning areas show a suggestive negative 
association with evaluative well-being, conditional on structural and 
temporal controls, but this effect is modest and does not extend to 
happiness, where no meaningful partisan gradient emerges.} Temporal shocks dominate the hedonic layer: happiness falls 
sharply during 2020--2022, whereas life satisfaction moves more modestly. 
{These patterns are robust across logistic and OLS 
specifications with clustered standard errors and align with well-being 
theory.} Interpreted as associations for the population of {
geolocated tweets}, the results show that large-scale, language-based 
indicators can help clarify why prior findings about the rural--urban divide may differ
by distinguishing the \emph{type} of well-being expressed, offering a 
transparent, reproducible complement to traditional surveys.
}

\maketitle

\clearpage

\section{Introduction}

Debates about a U.S.\ rural--urban divide typically foreground differences 
in education, values, opportunity structures, and political alignment. 
Since the 1990s these cleavages have acquired a pronounced electoral 
profile, with rural places trending more Republican and large urban areas 
more Democratic \citep{gimpel2020,gimpel2024,brown2024,brown2025,lin2024}. 
At the same time, research on subjective well-being (SWB) often reports 
that conservatives or Republicans express higher life satisfaction, though 
mechanisms and measurement are contested and results are sensitive to 
design choices 
\citep{napier2008,schlenker2012,burton2015,mandel2014,wojcik2015,
gimbrone2022}. Taken together, these strands suggest that geography and 
partisanship may jointly structure well-being in the United States, but 
existing evidence is mixed and often conflates distinct facets of SWB.

A first source of inconsistency is conceptual. \emph{Life satisfaction} 
and \emph{happiness} are related but distinct components of SWB: the 
former is an evaluative, cognitively integrated judgment about one's life, 
the latter a hedonic, momentary affective tone 
\citep{diener1984subjective,Kahneman1999,StevensonWolfers2008}. 
Conflating these dimensions can mask meaningful geographic and political 
patterns. A second source is empirical. Standard surveys rarely provide 
full county coverage or high-frequency measurement needed to trace both 
cross-sectional structure and temporal shocks, features that are crucial 
if well-being responds differently to slow-moving fundamentals (e.g., 
income, place) and fast-moving national events.

We address these challenges by constructing county-year indicators of 
\emph{life satisfaction} and \emph{happiness} from an archive of 2.6 
billion geolocated {tweets} (2013--2023) classified with 
a fine-tuned generative AI language model and linked to official county 
covariates (rural--urban continuum codes, household income, and election 
returns). We analyze 2014--2022 for which all covariates are available. 
Using precision-weighted logistic models for $P(\texttt{lifesat}>0)$ and 
$P(\texttt{happiness}>0)$ with year fixed effects and corroborating OLS 
specifications {with clustered standard errors}, we recover 
three robust results:

\begin{enumerate}[leftmargin=1.25em]

\item The rural--urban divide is \emph{two-sided}: rural counties exhibit 
systematically higher \emph{life satisfaction}, whereas urban and 
semi-urban counties exhibit higher \emph{happiness}. This inversion 
clarifies why prior U.S.\ studies reached mixed conclusions when 
aggregating across SWB dimensions.

\item {Income shows a positive but specification-sensitive 
association with life satisfaction (economically meaningful for \$10k 
increments in the logistic specification, but not robust in OLS with year 
fixed effects) and is negligible for happiness once other factors are held 
constant. Partisanship shows a suggestive negative association with life 
satisfaction (Democratic-leaning counties express lower evaluative 
well-being, conditional on controls), but this effect is modest, sensitive 
to the clustering correction, and does not extend to happiness. The 
rural $\times$ margin interaction is not statistically significant under 
clustered standard errors, though the negative pattern of marginal effects across 
rurality codes is visually suggestive of attenuation at both ends of 
the urban--rural continuum. This result should be treated as 
exploratory given the non-significance of the interaction term.}

\item Temporal shocks dominate the affective layer: happiness drops 
sharply in 2020--2022, while life satisfaction moves more mildly and 
heterogeneously, consistent with the evaluative--hedonic distinction 
\citep{helliwell2020world}.

\end{enumerate}

Methodologically, we show that social-media--based indicators can 
complement surveys at scale: they behave as theory predicts (i.e., life 
satisfaction and happiness are distinct yet correlated dimensions), respond 
plausibly to macro shocks, and reveal interpretable cross-sectional 
structure. {The robustness of the core spatial and temporal 
findings to clustered standard errors supports the validity of these 
indicators as measures of collective expressed well-being.} Substantively, 
the results help reconcile the U.S.\ ``puzzle'' by demonstrating that 
conclusions depend on \emph{which} facet of well-being is measured and on 
whether temporal dynamics and geographic context are modeled explicitly.

The paper is organized as follows: Section~\ref{sec:debate} reviews 
literatures on rural--urban and partisan cleavages and on the 
evaluative--hedonic distinction in SWB. Section~\ref{sec:data} describes 
the social-media indicators, weighting, and county covariates. 
Section~\ref{sec:lifestat_and_happiness_SM} documents the correlation 
structure between life satisfaction and happiness. 
Section~\ref{sec:analysis} models the determinants of \texttt{lifesat}; 
Section~\ref{sec:happiness} applies the same framework to \texttt{happiness} 
and highlights the urban affective advantage and pandemic-era declines. 
Section~\ref{sec:reconcile} compares full-model effects across outcomes to 
reconcile the puzzle. Section~\ref{sec:risks} discusses threats to 
inference and mitigations. Section~\ref{sec:conclusions} concludes with 
implications for research and place-based policy.

\section{The debate in the literature}\label{sec:debate}

Recent scholarship has emphasized that the rural--urban divide in the 
United States reflects not simply geographical but also 
social-psychological and partisan distinctions. In particular, this 
partisan and perhaps ideological divide also entails, according to 
several studies, a difference in terms of subjective well-being.

\subsection{Rural--Urban and Partisan Cleavages}

\citet{gimpel2024} theorize that satisfaction with one's place of 
residence and the psychological attachment to place are fundamental in 
structuring political attitudes along the rural--urban axis. Their 
findings show that population density and proximity to urban centers are 
associated with reduced place satisfaction, while discontent with one's 
environment predicts more progressive political preferences. Conversely, 
contentment with one's locale is linked to conservative viewpoints. These 
spatial divides persist even when individual-level characteristics (such 
as race, income, and education) are controlled for, suggesting a 
contextual effect of locality itself \citep{gimpel2020}. Two mechanisms 
are advanced: (1) physical and social distance between urban and rural 
populations discourages mutual understanding, and (2) the structuring of 
public opinion by population density leads to different political 
expressions in large cities versus small towns.

However, \citet{lin2024} question the primacy of spatial context, 
demonstrating that once partisanship and key demographics are accounted 
for, the apparent rural--urban rift in issue attitudes is largely an echo 
of national partisan divisions. Rural Democrats are no more conservative 
than urban Democrats on most issues, and likewise for Republicans, 
indicating that party affiliation outweighs place context for most policy 
positions, with only minor and residual contextual effects detected.

Historical and sociological perspectives \citep[as articulated 
by][]{brown2024,brown2025} further nuance this picture. The rural--urban 
political divide has developed through multi-faceted processes including 
economic restructuring, growing educational gaps, organizational 
influences (such as the role of evangelical churches), and racial 
dynamics. Notably, the partisan divergence across the rural--urban 
continuum is most pronounced among white Americans, while Black and Latino 
rural residents exhibit voting patterns more similar to their urban peers. 
Theories of ``linked fate'' and ``racialized social constraint'' suggest 
that people of color in rural areas are less likely to mobilize around 
rural identity, particularly given its contemporary associations with 
whiteness and exclusion.

\subsection{Partisan Differences in Life Satisfaction}

A parallel body of research has investigated how political orientation 
and specifically the Democrat--Republican divide relates to life 
satisfaction and subjective well-being in the U.S. population. 
\citet{napier2008} argue, via the system justification theory, that 
conservatives report higher levels of life satisfaction because they are 
more likely to rationalize social inequality and endorse meritocratic 
beliefs. Their multi-study evidence shows that this rationalization acts 
as a mediator: once controlled, the relationship between conservatism and 
well-being is reduced to non-significance. Moreover, increasing societal 
inequality exacerbates the happiness gap: liberals' happiness declines 
more sharply than conservatives' in periods of rising inequality.

\citet{schlenker2012} challenge the explanatory primacy of system 
justification, emphasizing instead conservative strengths in personal 
agency, religiosity, and positive outlook. Their empirical work indicates 
that these factors, more than system-justifying attitudes, mediate the 
higher life satisfaction found among conservatives. \citet{burton2015} 
further disentangle these dynamics, showing that personality traits, 
especially lower neuroticism and higher conscientiousness (as well as 
religiosity), explain away much of the association between conservatism 
and well-being.

Beyond individual traits, \citet{mandel2014} highlight the contextual 
sensitivity of Republicans' reported life satisfaction. Republican 
partisans not only identify more robustly with their party, but also tie 
their well-being more closely to the political climate, experiencing 
higher satisfaction when their party is in power. This ``favorability 
effect'' underscores how affective political orientation amplifies or 
dampens well-being according to external events.

Finally, \citet{wojcik2015} raise important methodological concerns. They 
reveal that much of the reported happiness advantage of conservatives 
emerges from self-report biases, particularly the tendency for 
conservatives to display higher self-enhancement. Objective indicators of 
positive affect (e.g., linguistic analysis, facial expressions, social 
media content) often paint a different picture, sometimes suggesting 
parity or even a liberal advantage in genuine emotional expression.

\citet{gimbrone2022} extend these findings to adolescents. Internalizing 
symptoms have risen most markedly among young liberal women, suggesting 
that the psychological costs of the current political climate are most 
acute in this group. Importantly, controlling for rural--urban status does 
not substantially affect these results, again pointing to ideology rather 
than place as the principal driver of well-being disparities.

Building on these premises, the relationship between the rural--urban 
divide and self-perceived well-being deserves more in-depth examination. 
The availability of new empirical evidence provides an opportunity to 
advance our understanding of this nexus, while also illuminating, to some 
extent, its implications for patterns of partisan polarization. 
{We note that the existing literature largely relies on 
survey-based, individual-level measures and does not systematically 
distinguish between evaluative and hedonic dimensions of well-being at 
the geographic level. Our contribution is to address this gap using 
large-scale social media data, while treating any partisan associations 
as contextual and potentially specification-sensitive, consistent with 
the mixed and contested evidence reviewed above.}

\subsection{Life Satisfaction and Happiness Are Not the Same Thing}

Subjective well-being (SWB) is commonly conceptualized as comprising at 
least two distinct components: a cognitive dimension, often referred to 
as \emph{life satisfaction}, and an affective dimension, typically 
described as \emph{happiness} or \emph{emotional well-being} 
\citep{diener1999subjective, feldman2010}. Life satisfaction 
reflects an individual's overall evaluative judgment about the quality of 
her/his own life as a whole, based on her/his own standards and 
aspirations. In contrast, happiness pertains to the balance of positive 
and negative short-run affect, i.e. how people feel in their day-to-day 
experiences \citep{diener1984subjective, stone2010well}.

Although life satisfaction and happiness are correlated, they are not 
interchangeable constructs. Happiness captures the affective tone of 
immediate experience, whereas life satisfaction is more reflective and 
evaluative, integrating past experiences and expectations for the future 
\citep{angner2010subjective}. As \citet{diener2009wellbeing} and others 
have argued, conflating these two dimensions can obscure meaningful 
variation across individuals, groups, or regions. For instance, the 
antecedents of affective well-being such as daily social interactions or 
health often differ from those influencing life satisfaction, such as 
income or goal attainment \citep{kahneman2006would, steptoe2015subjective}.

Recognizing the cognitive--affective distinction is crucial for 
interpreting well-being patterns across contexts. In particular, 
population-level studies increasingly demonstrate that life satisfaction 
and happiness respond differently to economic and social conditions 
\citep{helliwell2020world}. {The SWB literature also 
recognizes additional dimensions beyond this hedonic--evaluative 
dichotomy, including eudaimonic components such as meaning, purpose, and 
flourishing \citep{VanderWeele2017, diener2009wellbeing}. While 
these broader dimensions are not the focus of the present study, we 
acknowledge their importance and note that our data, drawn from the Human 
Flourishing Geographic Index, also encompasses indicators of meaning, 
purpose, and other eudaimonic domains \citep{iacus2025}.} Both dimensions 
should be measured and analyzed separately when assessing the geography 
or determinants of human well-being. {In our study, this 
duality emerges clearly and links systematically to the rural--urban 
dimension.} Similar evaluative--affective mismatches have been documented 
elsewhere. Using Gallup World Poll data, \cite{Greyling2025Cheerful} show 
that positive affect and life evaluation diverge substantially in 
Sub-Saharan Africa. Their results, although cross-national and 
survey-based, emphasize that affective and evaluative well-being can 
follow different patterns. Our rural--urban contrast is analogous in form 
and likewise becomes interpretable once the two dimensions and relevant 
covariates are modeled separately.

\section{The Data}\label{sec:data}
Life satisfaction and happiness social media indicators are part of a larger set of indicators from the project \textit{The Human Flourishing Geographic Index} (HFGI). 
The Human Flourishing Geographic Index dataset \citep{iacus2025} is conceptually inspired by Harvard’s Human Flourishing Program \citep{VanderWeele2017}, which defines flourishing as a multidimensional construct across six domains: happiness and life satisfaction, mental and physical health, meaning and purpose, character and virtue, close social relationships, and material and financial stability. Whereas the Global Flourishing Study provides cross-country, wave-based (only two waves are available at the time of this writing) measures from survey data, the HFGI dataset offers high-resolution indicators at county and state level for the United States. It has been obtained  analyzing approximately 2.6 billion geo-referenced tweets from the Harvard CGA Geotweet Archive \citep{Lewis16}, classified using a  fine-tuned large language model (LLM) \citep{finetuning2024}, to generate indicators corresponding to the Global Flourishing Study framework. The dataset spans January 2013 to June 2023 at both monthly and yearly frequency. 
The details on how these indicators have been computed through generative AI analysis are given in the Appendix. The interested reader can also find information on the fine-tuning strategy and the accuracy of the LLM classification in \cite{finetuning2024}. The  content of the HFGI dataset is illustrated in \cite{iacus2025}.

The social media indicators $\tt lifesat$ and $\tt happiness$ have been aggregated at county level. Their values vary in the interval $[-1,+1]$. As a measure of data quality, the dataset also includes the standard deviation of each statistics. These values will be used as weights in our models. Using these weights is a way to take into account the different levels of accuracy of the statistics due to different densities of tweets per geographical area.

In addition to the two flourishing indicators, we also add a set of control variables. We make use of the 2023 Rural-Urban Continuum Codes ($\tt rural$)  that distinguish U.S. metropolitan (metro, values 1-3) counties by the population size of their metro area, and non-metropolitan (nonmetro, values 4-9) counties by their degree of urbanization and adjacency to a metro area \citep{USDA_ERS_2024}. 

To control for county-level economic conditions, we include a measure of household income derived from the American Community Survey (ACS) 5-Year Estimates (\texttt{acs5}), which provides reliable small-area averages based on multi-year data \citep{uscb_acs5}.

To account for political partisanship, we include county-level presidential election results for the 2016 and 2020 U.S. elections, obtained from the MIT Election Data and Science Lab (MEDSL) repository \citep{medsl_presidential}.
County-level electoral results for U.S. midterm elections in 2014, 2018 and 2022, were obtained from the OpenElections project \citep{openelections}, which provides standardized, open-access election data compiled from official state sources. From these data, we compute the vote \texttt{margin} as the difference between the Democratic and Republican vote shares, expressed in percentage points. This standard definition follows the convention in political science for measuring electoral competitiveness \citep{ansolabehere2002incumbency, fowler2008political}.

To control for unobserved heterogeneity over time, we incorporate year fixed effects, capturing global shocks (e.g., the pandemic crisis), policy shifts, or macroeconomic conditions that may simultaneously affect all counties. This  follows the standard panel data practices in applied econometrics \citep{wooldridge2010econometric, angrist2009mostly}.

\section{Life Satisfaction and Happiness on Social 
Media}\label{sec:lifestat_and_happiness_SM}

Table~\ref{tab:corr} shows that the correlation between Twitter-based 
life satisfaction and happiness is strong but not perfect, and it 
decreases over time. This pattern suggests that these two affective 
indicators, while conceptually related, capture distinct facets of 
well-being as expressed in online discourse.

\begin{table}[!ht]
\caption{Correlation between Life Satisfaction and Happiness by Year}
\label{tab:corr}
\centering
\begin{tabular}[t]{rrll}
\toprule
Year & N & Correlation (95\% CI) & p-value\\
\midrule
2014 & 3143 & 0.805 [0.792, 0.817] & $< 10^{-300}$ \\
2016 & 3140 & 0.496 [0.469, 0.522] & $7.5 \times 10^{-195}$ \\
2018 & 3143 & 0.605 [0.582, 0.627] & $3.5 \times 10^{-313}$ \\
2020 & 3143 & 0.558 [0.533, 0.582] & $1.2 \times 10^{-256}$ \\
2022 & 3140 & 0.442 [0.413, 0.470] & $3.5 \times 10^{-150}$ \\
\bottomrule
\end{tabular}
\vspace{1ex}
\begin{flushleft}
\footnotesize
\textit{Note:} All correlations are statistically significant at 
$p < 10^{-150}$. Values in brackets indicate 95\% confidence intervals.
\end{flushleft}
\end{table}

{Figure~\ref{fig:fig0} provides a first descriptive look 
at the raw, unadjusted association between rurality and the two well-being 
dimensions. County-level means of \texttt{lifesat} and \texttt{happiness} 
are plotted across the nine rurality codes (1--9) without any controls. 
The two series display opposite gradients: life satisfaction increases 
monotonically from urban (code~1) to rural (code~9) counties, while 
happiness decreases along the same continuum. This unadjusted pattern 
already suggests the rural--urban paradox that the subsequent models 
formalize and explain.}

{\begin{figure}[!ht]
    \centering
    \includegraphics[width=0.75\linewidth]{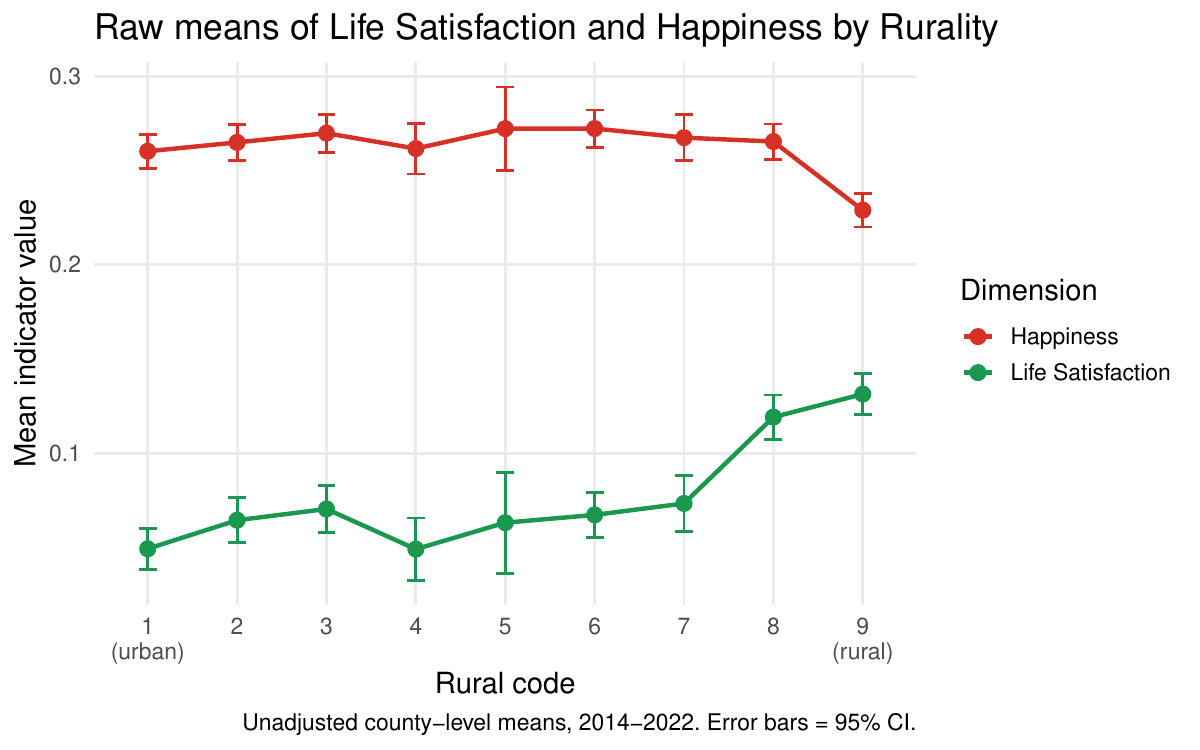}
    \caption{Unadjusted county-level means of \texttt{lifesat} and 
\texttt{happiness} by rurality code (1--9), averaged across 2014--2022. 
Error bars denote 95\% confidence intervals. Life satisfaction increases 
monotonically with rurality while happiness decreases, documenting the 
raw rural--urban divergence in expressed well-being prior to any 
covariate adjustment.}
    \label{fig:fig0}
\end{figure}}

\begin{figure}[!ht]
    \centering
    \includegraphics[width=0.75\linewidth]{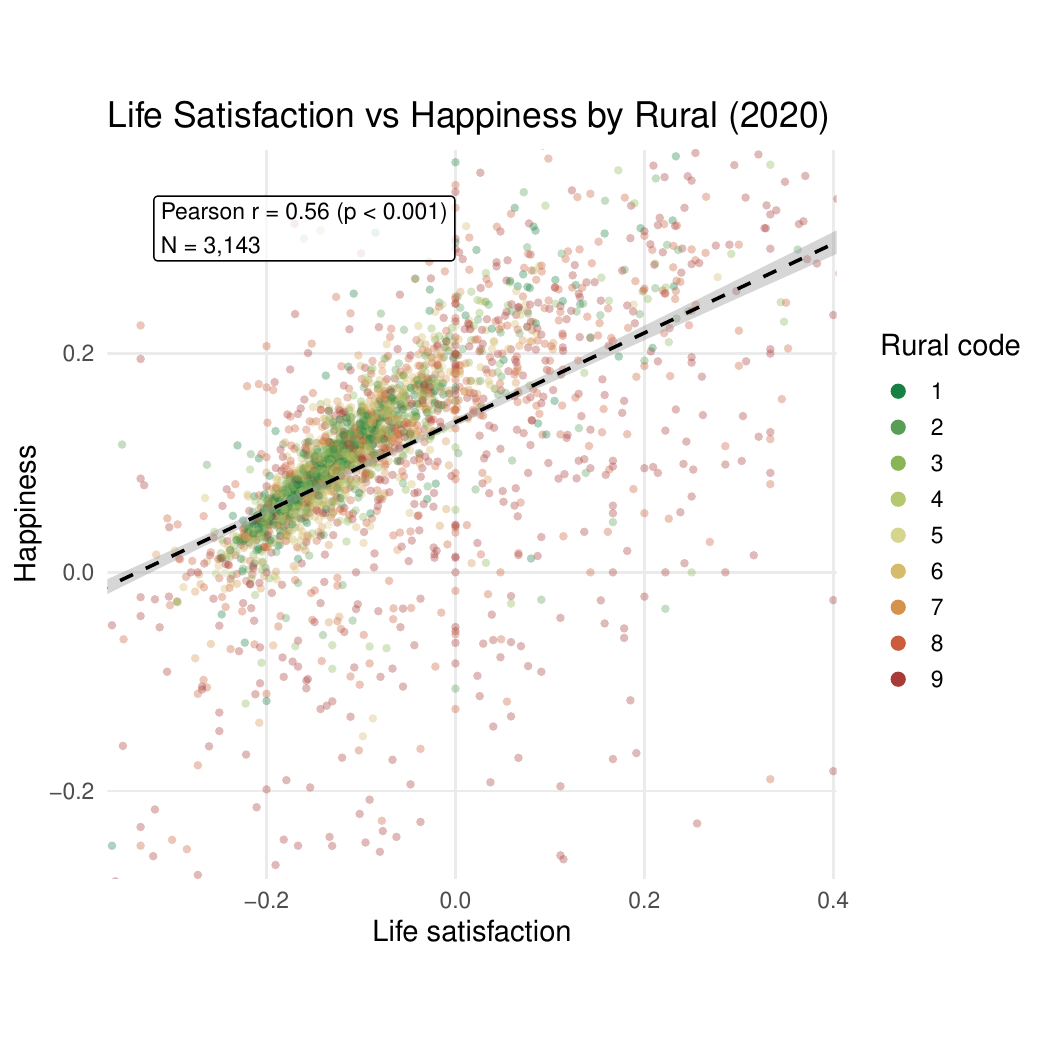}
    \caption{Relationship between Twitter-based life satisfaction and 
happiness across U.S. counties in 2020. Each point represents a county, 
colored by its rurality code (1--9) {using a diverging 
RdYlGn color scale, from green (urban, code 1) to red (most remote rural, 
code 9)}. The dashed line shows the fitted linear regression with 95\% 
confidence band. The correlation is moderate and positive ($r = 0.55$, 
$p < 0.001$, $N = 3{,}143$), indicating that, while the two affective 
indicators are related, they capture distinct facets of expressed 
well-being.}
    \label{fig:fig1}
\end{figure}

\begin{figure}[!ht]
    \centering
    \includegraphics[width=0.95\linewidth]{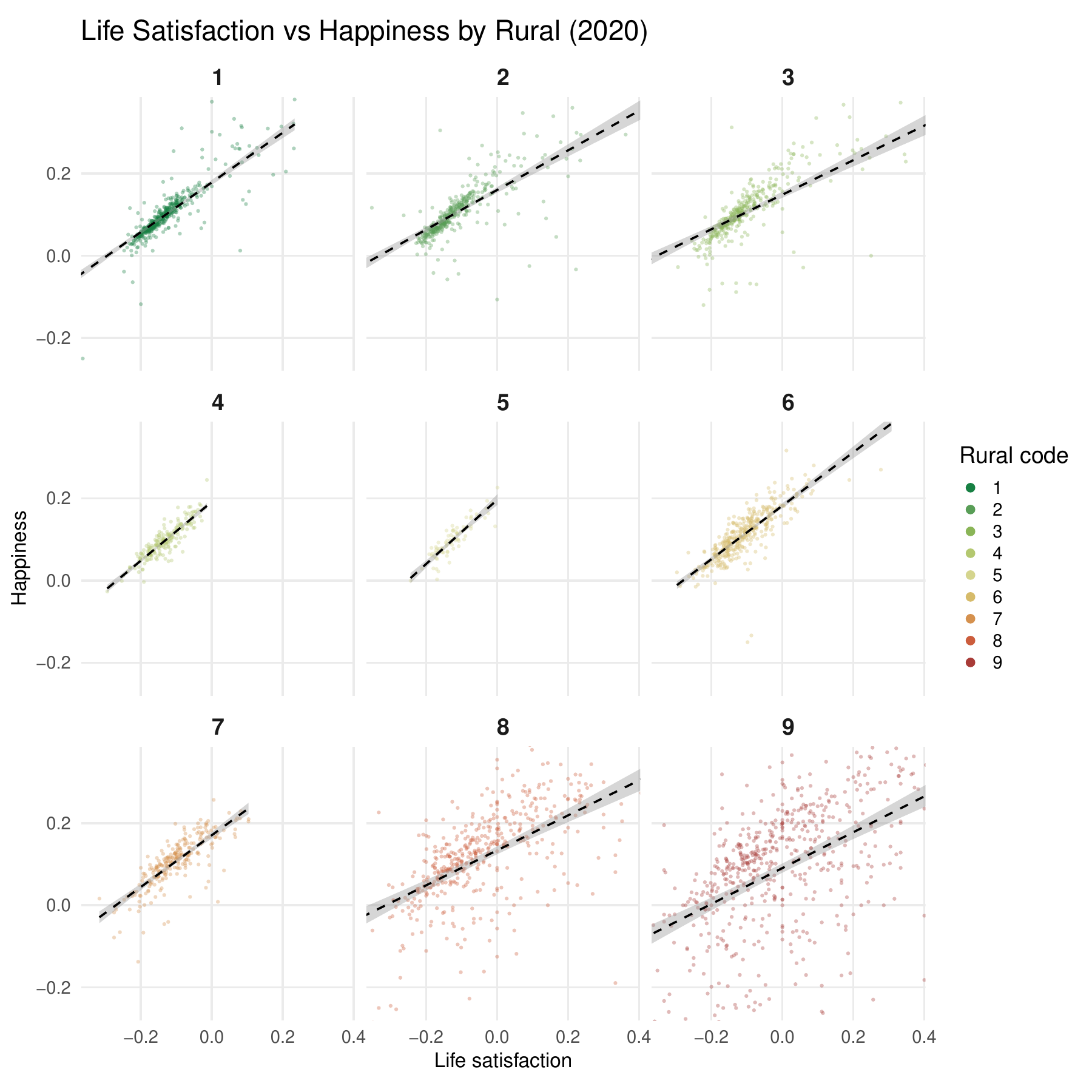}
    \caption{Relationship between Twitter-based life satisfaction and 
happiness by rurality code (1--9) in 2020. Each panel shows counties 
within the same rural classification, with a fitted linear regression 
(dashed line) and 95\% confidence band. {Colors follow 
the same RdYlGn diverging scale as Figure~\ref{fig:fig1}.} The positive 
association between life satisfaction and happiness is consistent across 
all levels of rurality, though the strength of the correlation varies by 
group. Notably, panels for codes 8 and 9 display an almost quadratic 
pattern, reflecting greater heterogeneity in expressed well-being among 
the most remote rural counties.}
    \label{fig:fig2}
\end{figure}

Figures~\ref{fig:fig1} and~\ref{fig:fig2} illustrate, as an example, 
the relationship between our Twitter-based indicators of life satisfaction 
and happiness for the year 2020. Figure~\ref{fig:fig1} shows a clear 
positive association across all U.S. counties ($r = 0.55$, $p < 0.001$), 
indicating that higher life satisfaction tends to coincide with higher 
happiness in online expressions of well-being. Figure~\ref{fig:fig2} 
disaggregates this relationship by rurality code (1--9). The association 
is consistently positive across all county types, but its strength is 
notably higher in urban and semi-urban areas (low rural codes) than in 
predominantly rural counties. This suggests that in more urban contexts, 
expressions of happiness and life satisfaction tend to co-move more 
closely, whereas in rural settings these dimensions appear more weakly 
coupled, reflecting potentially distinct underlying drivers of well-being. 
In the next two sections we will analyze separately the determinants of 
those two dimensions of well-being, while in 
Section~\ref{sec:reconcile} we will try to reconcile the evidence.

\section{Modeling the Determinants of Life Satisfaction}\label{sec:analysis}

We estimate a sequence of nested models to identify the structural and 
contextual determinants of Twitter-based life satisfaction. 
Model~1 includes three core predictors: rurality (\texttt{rural}), partisan 
vote margin (\texttt{margin}, defined as Democratic minus Republican share), 
and the short-term affective indicator (\texttt{happiness}). 
Model~2 adds median household income from the U.S. Census Bureau's American 
Community Survey 5-Year Estimates (\texttt{acs5}), capturing county-level 
economic well-being. 
Model~3 incorporates year fixed effects to account for national shocks and 
time-specific factors affecting all counties. 
Finally, Model~4 includes an interaction between \texttt{rural} and 
\texttt{margin}, allowing the association between partisanship and life 
satisfaction to differ between rural and urban areas.

{We model the probability that county-level life satisfaction 
is above zero using a logistic specification, rather than a linear regression. 
The dependent variable $\texttt{lifesat} \in [-1,+1]$ is bounded and our 
classification scheme assigns scores of $-1$ (negative), $+0.5$ (somewhat 
positive), and $+1$ (positive), with $0$ reserved for tweets in which the 
dimension is not expressed. The threshold at zero therefore corresponds to the natural boundary between 
negative and positive expressed sentiment, consistent with the polarity 
structure of the underlying construct.
}

Because life satisfaction and happiness are closely related yet conceptually 
distinct---happiness reflecting short-term affective well-being and life 
satisfaction representing a longer-term cognitive evaluation---including 
\texttt{happiness} as a covariate in the model for \texttt{lifesat} allows 
us to control for the affective baseline and assess whether socioeconomic and 
contextual factors retain independent effects. Formally, the general model 
estimated across counties $c$ and years $t$ can be written as:
\begin{align}
\label{eq:fullmodel}
\text{logit}\,\Pr(Y_{ct}=1) =
\beta_0
&+ \beta_1  \text{Happiness}_{ct}
+ \beta_2 \text{Rural}_c
+ \beta_3 \text{Margin}_{ct}
+ \beta_4 \text{Income}_c \notag \\
&+ \beta_5 (\text{Rural}_c \times \text{Margin}_{ct})
+ \gamma_t
+ \epsilon_{ct},
\end{align}
where $Y_{ct} = 1$ indicates whether county $c$ in year $t$ exhibits a 
positive life-satisfaction indicator (${\texttt{lifesat}_{ct} > 0}$), 
$\gamma_t$ denotes year fixed effects and $\epsilon_{ct}$ an idiosyncratic 
error term. Table~\ref{tab:lifesat} reports the results.

All models are estimated with weights equal to the inverse of the standard 
deviation of the county's life satisfaction estimate 
($1/\texttt{lifesat\_se}^2$), giving greater weight to counties where the 
dependent variable is more precisely measured. 

{These precision weights serve a dual purpose. First, they 
down-weight counties with sparse tweet coverage, where the aggregate 
sentiment signal is noisier. Second, the standard deviation of the 
LLM-based estimate is strongly correlated with county population on a 
log scale ($r = 0.70$, $\log(1/\texttt{lifesat\_se}^2)$ vs.\ 
$\log(\text{population})$), confirming that precision weights effectively 
capture population size while also incorporating classification uncertainty 
from the LLM classification. Weighting by raw population alone would ignore this 
additional source of noise. We therefore prefer precision weights over 
population weights as a more informative and theoretically grounded measure 
of estimate reliability.}

{As a robustness check, we also estimate OLS models on the 
continuous \texttt{lifesat} scale, reported in Table~\ref{tab:lifesatOLS}. 
The OLS specifications allow us to verify that the main findings are not 
artifacts of the logistic dichotomization and to assess the sensitivity 
of results to the choice of model form. Coefficients and signs are 
substantively consistent across logistic and OLS specifications, 
supporting the robustness of the results.}

{Since the panel structure of the data includes the same 
counties observed across multiple years, standard errors from both the 
logistic and OLS models may understate uncertainty due to within-county 
serial correlation. We therefore report heteroskedasticity-robust standard 
errors clustered at the county level (HC1) for all specifications, which are 
asymptotically valid under within-county correlation of arbitrary form 
\citep{wooldridge2010econometric}. For the logistic models, average marginal effects 
with clustered standard errors are computed via the \texttt{marginaleffects} 
package \citep{ArelBundock2024}.
 The empirical 
results are discussed in the next section.}

\begin{table}[ht]
\centering
\caption{Average Marginal Effects for the full 
\texttt{lifesat} model in equation~\eqref{eq:fullmodel}. 
Clustered standard errors (HC1) by county via 
\texttt{marginaleffects} \citep{ArelBundock2024}.} 
\label{tab:lifesat_margins}
\begin{tabular}{lrrrlrr}
  \toprule
Predictor & AME & SE & z & p & lower & upper \\ 
  \midrule
acs5 ($\times$ \$10,000) & 0.0054 & 0.0019 & 2.88 & 0.004 & 0.0017 & 0.0091 \\ 
  happiness & 1.7623 & 0.0339 & 52.01 & $<$0.001 & 1.6959 & 1.8287 \\ 
  margin & -0.0262 & 0.0100 & -2.61 & 0.009 & -0.0459 & -0.0065 \\ 
  rural & 0.0140 & 0.0015 & 9.54 & $<$0.001 & 0.0111 & 0.0168 \\ 
  year = 2016 & 0.2302 & 0.0881 & 2.61 & 0.009 & 0.0575 & 0.4029 \\ 
  year = 2018 & 0.0257 & 0.0113 & 2.27 & 0.023 & 0.0035 & 0.0479 \\ 
  year = 2020 & 0.0801 & 0.0093 & 8.66 & $<$0.001 & 0.0620 & 0.0983 \\ 
  year = 2022 & 0.1206 & 0.0120 & 10.08 & $<$0.001 & 0.0972 & 0.1441 \\ 
   \bottomrule
\end{tabular}
\end{table}

\subsection{Empirical Results and Interpretation}

The results in Table~\ref{tab:lifesat} and Table~\ref{tab:lifesat_margins} 
show that short-term \texttt{happiness} is by far the strongest predictor 
of \texttt{lifesat} on Twitter (AME $=1.762$, $p<0.001$). Counties with 
higher average happiness also display substantially greater probabilities 
of expressing positive life satisfaction. This confirms the tight link 
between affective and evaluative well-being, while leaving room for 
meaningful spatial and temporal structure captured by other covariates.
Model fit improves markedly when introducing year fixed effects (AIC drops 
from 47{,}653 in Model~2 to 42{,}017 in Model~3; 
Table~\ref{tab:lifesat}), underscoring the importance of national shocks.

Rurality (\texttt{rural}) is positively associated with life satisfaction 
across all model specifications. Its marginal effect (AME $=0.014$, 
$p<0.001$) indicates that, holding other covariates constant, more-rural 
counties have higher probabilities of expressing positive life satisfaction. 
This relationship remains highly significant even after adjusting for 
income, political margin, and time effects, underscoring the persistence 
of rural well-being advantages in social media expression.

{Household income (\texttt{acs5}) is positively associated 
with life satisfaction in the logistic specification (AME $= 0.005$ per 
\$10{,}000, $p=0.004$; Table~\ref{tab:lifesat_margins}). Interpreting the 
logit scale (coefficient $\approx 0.120$ per \$10{,}000), a \$10{,}000 
increase maps to about a $e^{0.12} \approx 1.13$ (13\%) increase in the 
\emph{odds} of $\texttt{lifesat}>0$. However, this effect is not 
statistically distinguishable from zero in the OLS specification once year 
fixed effects are included (Table~\ref{tab:lifesatOLS}), suggesting that 
the income gradient is modest and sensitive to model form. We therefore 
interpret the income effect as indicative rather than conclusive: economic 
prosperity may contribute incrementally to long-term evaluative well-being, 
but this association is not robust across all specifications.}

{The partisan vote margin (\texttt{margin}) has a negative 
marginal effect (AME $=-0.026$, $p=0.009$), indicating that, conditional 
on other covariates, Democratic-leaning counties tend to show slightly 
lower life satisfaction than Republican-leaning ones. This effect, while 
statistically significant, is modest in magnitude and should be interpreted 
with caution: \texttt{margin} is nearly time-invariant at the county 
level (most of its variation is cross-sectional rather than 
within-county over time)  and its coefficient is sensitive to the 
clustering correction. We therefore treat this as 
suggestive rather than conclusive evidence of a partisan gradient in 
evaluative well-being.}

Year fixed effects are large and positive, especially in 2016 and 2022 
(AMEs $\approx 0.23$ and $0.12$, respectively; 
Table~\ref{tab:lifesat_margins}), capturing pronounced nationwide shifts 
in expressed life satisfaction. These temporal swings dominate much of 
the overall variation in expressed well-being, consistent with the 
influence of major national events and economic or political cycles on 
collective sentiment.

{The interaction between partisanship and rurality 
(\texttt{margin:rural}) is not statistically significant under clustered 
standard errors (Table~\ref{tab:lifesat}), consistent with the 
near-collinearity between \texttt{margin} and \texttt{rural} discussed 
above. Nevertheless, the average marginal effect of \texttt{margin} 
computed separately for each rurality code (Figure~\ref{fig:margin_rural}; 
Table~\ref{tab:margin_by_rural} in the Appendix) is visually consistent 
with a negative gradient across all nine codes, with the effect 
{appearing} somewhat stronger in semi-rural counties 
(codes 6--8, AME $\approx -0.032$ to $-0.036$) and borderline 
non-significant at the extremes (codes 1 and 9, $p \approx 0.067$). 
This pattern is {visually suggestive of} attenuation at 
both ends of the urban--rural continuum, though the interaction term is 
not statistically significant and this interpretation should be treated 
as exploratory.}

\noindent
Taken together, three key findings emerge:
\textit{i)} Short-term happiness is the dominant affective determinant of 
life satisfaction, yet structural and contextual factors retain significant 
marginal effects once affect is controlled for; 
\textit{ii)} rural counties consistently exhibit higher life satisfaction 
probabilities, {while income and political margin show 
suggestive but specification-sensitive associations that should be 
interpreted with caution;}
\textit{iii)} temporal dynamics are sizable: year AMEs are strongly 
positive for 2016, 2018, 2020, and 2022 
(Table~\ref{tab:lifesat_margins}), reflecting nationwide shifts in 
expressed evaluations beyond local covariates.

\subsection{Visual Representation of the Results}

To illustrate how income, partisanship, and rurality jointly shape life 
satisfaction, we present model-predicted probabilities for 2022.

Figure~\ref{fig:fig4} plots the predicted probability of expressing
above-zero life satisfaction, $P(\texttt{lifesat}>0)$,
across the Democratic–Republican vote margin and three income strata
(20th, 50th, and 80th percentiles of median household income).
Across all panels, two patterns emerge.
First, rural counties (high rural codes) display higher baseline 
probabilities of positive life satisfaction than urban ones,
consistent with the positive main effect of \texttt{rural}.
Second, life satisfaction declines as the Democratic vote share increases. 
The plotted lines are visually suggestive of a somewhat steeper decline in 
intermediate-to-more-rural counties, although the interaction term 
\texttt{margin:rural} is not statistically significant under clustered 
standard errors and should not be interpreted as a confirmed effect. 
Figure~\ref{fig:margin_rural} displays the average marginal effect of 
\texttt{margin} by rurality code; the pattern is visible, but especially 
toward the extremes it remains statistically uncertain.

{\begin{figure}[!ht]
    \centering
    \includegraphics[width=0.85\linewidth]{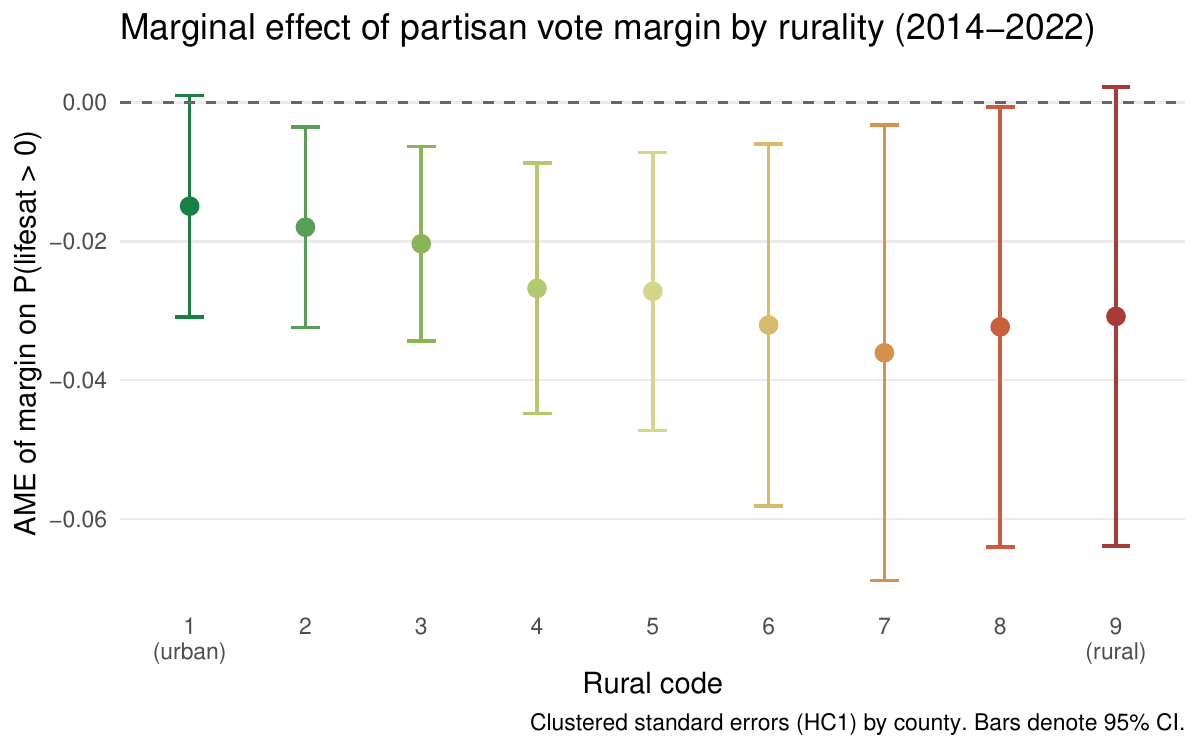}
\caption{{Marginal effect of partisan vote margin (\texttt{margin}) on 
$\Pr(\texttt{lifesat} > 0)$ by rurality code, from the full model 
(Model~4). Points represent average marginal effects; bars denote 95\% 
confidence intervals with clustered standard errors (HC1) by county. 
The effect is consistently negative across all rurality codes, 
indicating that Democratic-leaning counties display lower life 
satisfaction regardless of urban--rural context, though the effect {appears} strongest in semi-rural 
counties (codes 6--8) and borderline non-significant at the extremes 
(codes 1 and 9).}}
    \label{fig:margin_rural}
\end{figure}}

Figure~\ref{fig:fig5} explores the moderating role of income.
Here, the predicted probability of positive life satisfaction
is shown as a function of median household income, 
with separate panels for Republican-leaning, competitive, 
and Democratic-leaning counties.
Across all partisan contexts, life satisfaction rises monotonically 
with income, confirming the positive income coefficient in 
Table~\ref{tab:lifesat}.
Moreover, the rural–urban gap widens at higher income levels:
high-income rural counties show the largest advantage,
whereas low-income counties converge across rurality codes.
This pattern suggests that material prosperity amplifies
spatial and political differences in evaluative well-being.

Although Figures~\ref{fig:fig4} and~\ref{fig:fig5} display the results 
for 2022, similar patterns are observed across all years,
underscoring the robustness of these gradients in subjective life 
satisfaction.

\begin{figure}[!ht]
    \centering
    \includegraphics[width=0.85\linewidth]{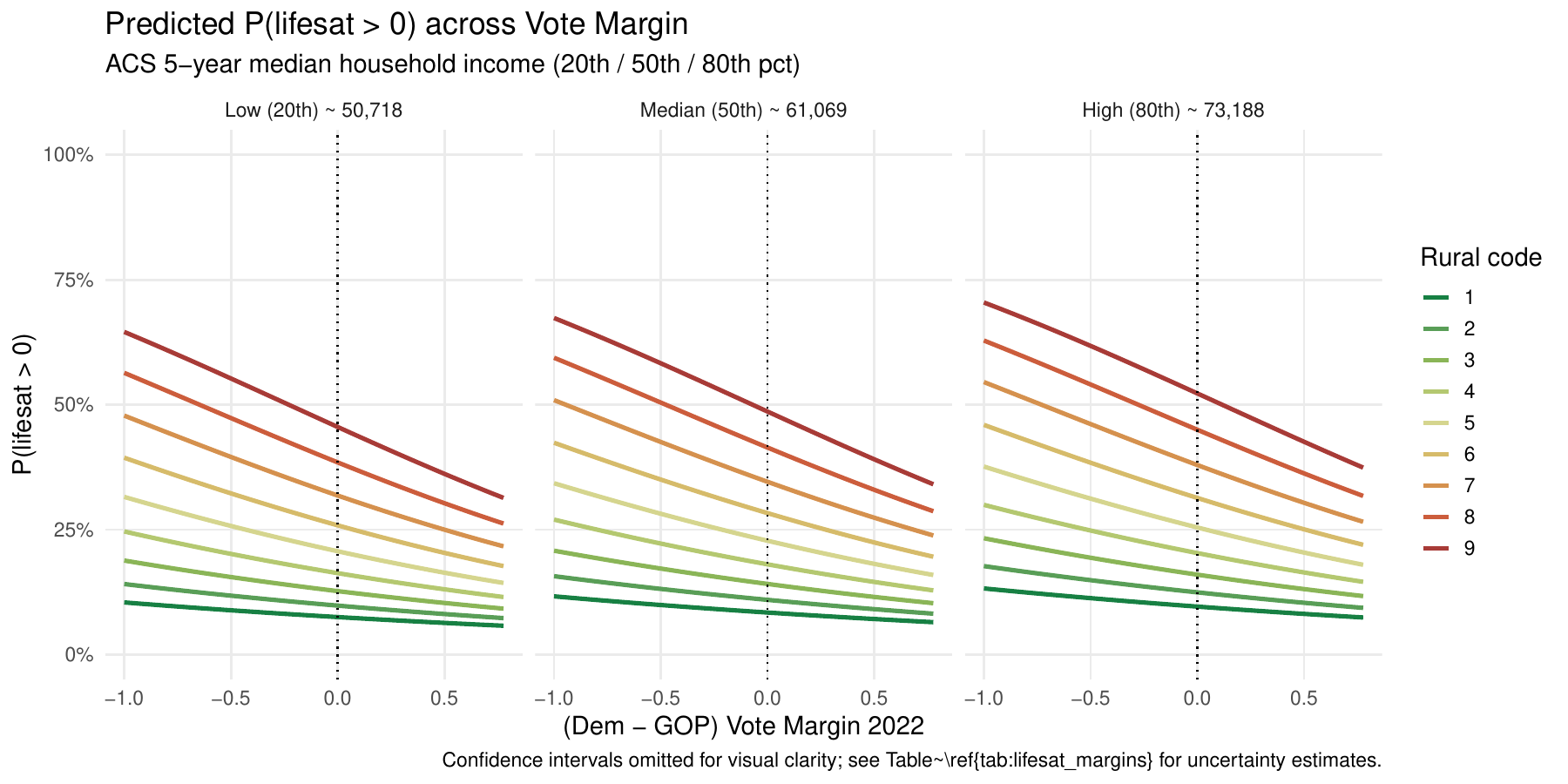}
\caption{Predicted probability of expressing positive life satisfaction 
($P(\texttt{lifesat}>0)$) across the Democratic--Republican vote margin 
for U.S.\ counties in 2022, based on the full model in 
equation~\eqref{eq:fullmodel}. Panels correspond to counties at the 
20th, 50th, and 80th percentiles of median household income (ACS 5-year 
estimates). Lines represent rurality codes (1--9). Life satisfaction 
decreases as the Democratic vote share increases, with a 
{visually suggestive steeper decline in more-rural 
counties (higher rural codes); note however that the 
\texttt{margin:rural} interaction term is not statistically significant 
under clustered standard errors (see Figure~\ref{fig:margin_rural}). 
Confidence intervals are omitted for visual clarity given the overlap 
across nine rurality codes; uncertainty estimates are reported in 
Table~\ref{tab:lifesat_margins}.}}
    \label{fig:fig4}
\end{figure}

\begin{figure}[!ht]
    \centering
    \includegraphics[width=0.85\linewidth]{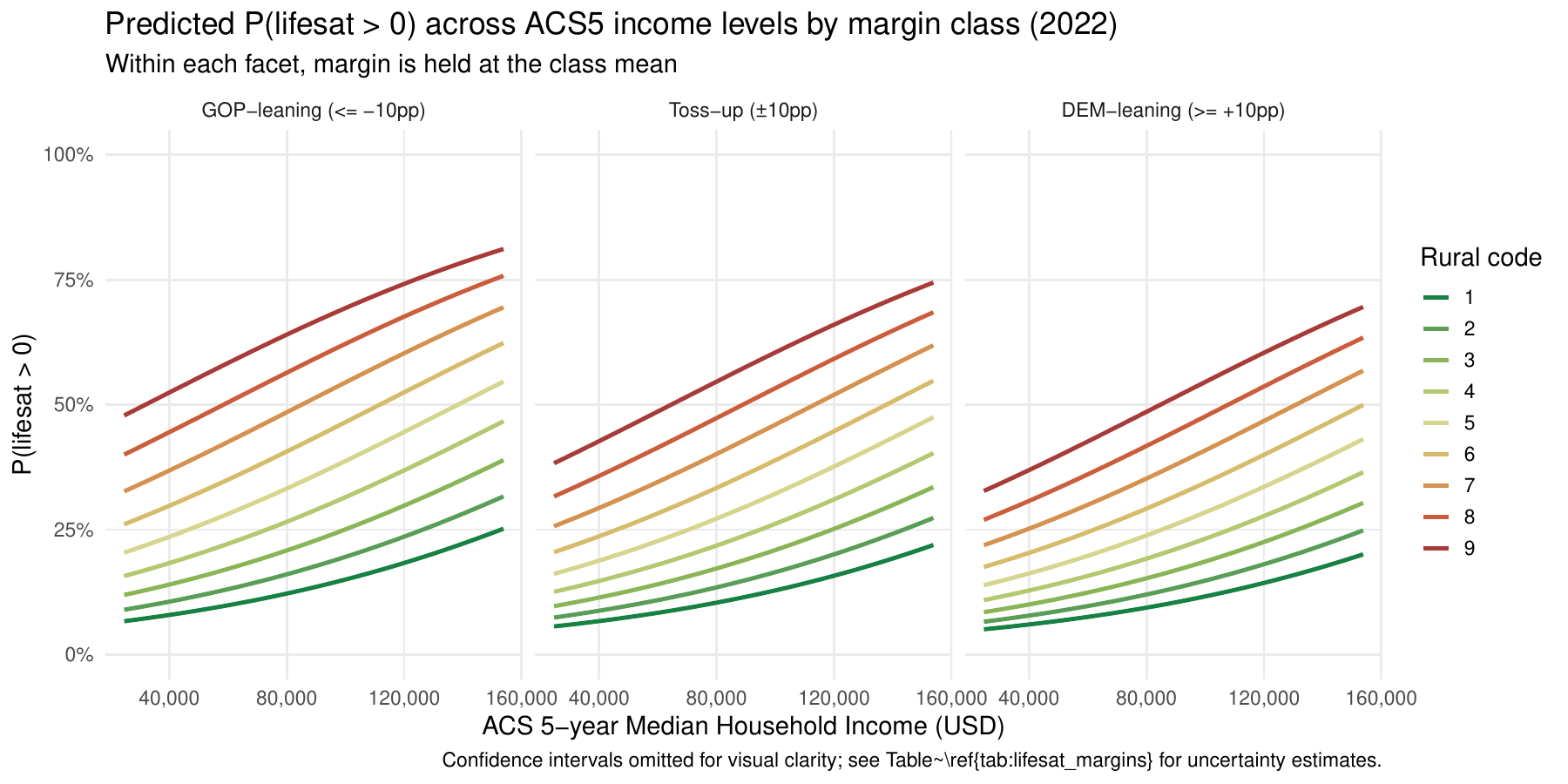}
\caption{Predicted probability of expressing positive life satisfaction 
($P(\texttt{lifesat}>0)$) across median household income levels for 
U.S.\ counties in 2022, based on the full model in 
equation~\eqref{eq:fullmodel}. Panels correspond to 
Republican-leaning ($\leq -10$ pp), toss-up ($\pm10$ pp), and 
Democratic-leaning ($\geq +10$ pp) counties. Within each panel the 
partisan margin is held at the class mean. Lines represent rurality 
codes (1--9). Life satisfaction increases with household income across 
all partisan contexts, and the rural--urban gap widens at higher 
incomes, suggesting that material prosperity amplifies spatial 
differences in evaluative well-being. {Confidence 
intervals are omitted for visual clarity given the overlap across nine 
rurality codes; uncertainty estimates are reported in 
Table~\ref{tab:lifesat_margins}.}}
    \label{fig:fig5}
\end{figure}

\newpage
\eject

\section{Modeling the Determinants of Happiness}\label{sec:happiness}

We now turn to the affective dimension of well-being and model the 
determinants of the \texttt{happiness} indicator, providing a direct 
complement to the life-satisfaction analysis.

As in the previous section, we estimate a sequence of weighted logistic 
regression models in which the dependent variable is a binary indicator 
equal to~1 when a county's average happiness score is positive 
($\texttt{happiness} > 0$) and~0 otherwise. {The threshold 
at zero corresponds to the natural boundary between negative and positive 
expressed affective tone, consistent with the polarity structure of the 
underlying construct, as discussed in Section~\ref{sec:analysis}.}
Each model is estimated using inverse-variance weights 
($1/\texttt{happiness\_se}^2$), giving greater weight to counties where 
the happiness measure is estimated more precisely. {As in the life satisfaction models, these precision weights 
are strongly correlated with county population on a log scale 
($r = 0.75$, $\log(1/\texttt{happiness\_se}^2)$ vs.\ 
$\log(\text{population})$), and therefore serve as an effective proxy for 
population size while also incorporating classification uncertainty from 
the LLM classification.}
Formally, the general model estimated across counties $c$ and years $t$ 
can be written as:
\begin{align}
\label{eq:happiness_full}
\text{logit}\,\Pr(Y_{ct} = 1) =
\beta_0
&+ \beta_1\,\text{LifeSat}_{ct} + \beta_2\,\text{Rural}_c
+ \beta_3\,\text{Margin}_{ct} \notag \\
&+ \beta_4\,\text{Income}_c 
+ \beta_5\,(\text{Rural}_c \times \text{Margin}_{ct}) \notag\\
&+ \gamma_t + \epsilon_{ct},
\end{align}
where $Y_{ct} = 1$ indicates whether county $c$ in year $t$ exhibits a 
positive happiness indicator (${\texttt{happiness}_{ct} > 0}$), $\gamma_t$ 
captures year fixed effects and $\epsilon_{ct}$ is an idiosyncratic error 
term. As before, the dependent variable is bounded in $[-1,+1]$; we 
therefore model the probability that county-level happiness is positive 
using a logistic link, focusing on the directional (positive vs.~negative) 
aspect of affective tone.

{As a robustness check, we also estimate OLS models on the 
continuous \texttt{happiness} scale, reported in 
Table~\ref{tab:happinessOLS}. Coefficients and signs are substantively 
consistent across logistic and OLS specifications. For both, 
heteroskedasticity-robust standard errors clustered at the county level 
(HC1) are reported throughout, following the same approach as in 
Section~\ref{sec:analysis}.}

\subsection{Empirical Results and Interpretation}

Table~\ref{tab:happiness} and Table~\ref{tab:happiness_margins} summarize 
the results of the weighted logistic regressions, parallel to the 
life-satisfaction models. The inclusion of life satisfaction as a predictor 
substantially improves model fit ($\Delta$Deviance~=~3{,}001, $p < 0.001$), 
confirming a strong association between long-term evaluative well-being and 
short-term affective tone. However, even after accounting for life 
satisfaction, contextual and temporal variables remain highly significant, 
indicating that happiness responds to distinct structural and temporal 
factors.

\begin{table}[ht]
\centering
\caption{Average Marginal Effects for the full 
\texttt{happiness} model. Clustered standard errors (HC1) by county 
via \texttt{marginaleffects} \citep{ArelBundock2024}.} 
\label{tab:happiness_margins}
\begin{tabular}{lrrrlrr}
  \toprule
Predictor & AME & SE & z & p & lower & upper \\ 
  \midrule
acs5 ($\times$ \$10,000) & -0.0001 & 0.0007 & -0.08 & 0.936 & -0.0014 & 0.0013 \\ 
  lifesat & 0.1409 & 0.0330 & 4.27 & $<$0.001 & 0.0762 & 0.2057 \\ 
  margin & -0.0020 & 0.0096 & -0.21 & 0.831 & -0.0208 & 0.0167 \\ 
  rural & -0.0066 & 0.0017 & -3.95 & $<$0.001 & -0.0099 & -0.0033 \\ 
  year = 2016 & -0.0703 & 0.0247 & -2.84 & 0.004 & -0.1187 & -0.0219 \\ 
  year = 2018 & -0.0062 & 0.0022 & -2.77 & 0.006 & -0.0106 & -0.0018 \\ 
  year = 2020 & -0.0196 & 0.0039 & -5.06 & $<$0.001 & -0.0272 & -0.0120 \\ 
  year = 2022 & -0.0406 & 0.0145 & -2.80 & 0.005 & -0.0689 & -0.0122 \\ 
   \bottomrule
\end{tabular}
\end{table}

Life satisfaction is the dominant predictor of happiness 
(AME~=~0.14, $p < 0.001$), consistent with the strong correlation 
observed in Table~\ref{tab:corr}.

Yet several contrasts emerge relative to the life satisfaction model.

Rurality (\texttt{rural}) has a significant \emph{negative} effect 
(AME~=~$-$0.007, $p < 0.001$): holding other variables constant, 
more-rural counties have lower predicted probabilities of positive 
happiness. This result is the mirror image of the life satisfaction model, 
indicating an \emph{urban affective advantage} in social media expressions 
of well-being.

{The partisan vote margin (\texttt{margin}) has no 
statistically significant effect on happiness (AME~=~$-$0.002, $p=0.831$), 
indicating that, once structural and temporal factors are accounted for, 
Democratic and Republican-leaning counties do not differ meaningfully in 
their expressed affective tone. This stands in contrast to the 
life satisfaction model, where a suggestive negative effect of 
\texttt{margin} was detected. The interaction between partisanship and 
rurality (\texttt{margin:rural}) is likewise not statistically significant 
under clustered standard errors (Table~\ref{tab:happiness}), suggesting 
that the potential partisan gradient in well-being is specific to the evaluative 
dimension and does not extend to hedonic affect.}

Household income (\texttt{acs5}) has no statistically significant effect 
once other factors are controlled (AME $\approx 0$, $p=0.936$). 
{This null result is consistent across both logistic and OLS 
specifications with year fixed effects, and mirrors the pattern observed 
for life satisfaction: the income gradient in well-being, to the extent it 
exists, operates primarily through the evaluative rather than the hedonic 
dimension.}

Year fixed effects are large and negative, with steep drops in 2020 and 
2022 (AMEs~=~$-$0.020 and $-$0.041), capturing the national decline in 
affective tone during the pandemic and its aftermath. These shifts align 
with evidence from traditional surveys documenting widespread emotional 
deterioration in the same period \citep{Deaton2013,Sibley2020}.

{Note that the sign of the year fixed effects differs 
markedly from the life satisfaction model, where year AMEs were large and 
\emph{positive}. This divergence is theoretically informative: happiness, 
as a hedonic and momentary indicator, is highly sensitive to national mood 
shocks (COVID-19 pandemic, political turbulence), whereas life satisfaction, as a slower-moving evaluative construct, 
shows more modest temporal variation and, notably, in the positive 
direction, in contrast to the sharp negative swings observed for 
happiness.}

Taken together, these findings confirm that happiness and life satisfaction, 
while correlated, are shaped by different forces. Happiness is more 
sensitive to temporal shocks and to the social expressiveness of urban 
environments, whereas life satisfaction is driven more by structural and 
evaluative factors such as income, rurality, and long-term stability.

{In the OLS specification (Table~\ref{tab:happinessOLS}), 
all main results are confirmed: \texttt{rural} remains negative and 
significant across all models, \texttt{margin} and \texttt{margin:rural} 
are not significant with year fixed effects, and year effects are large 
and negative for 2020 and 2022. One notable feature is that \texttt{acs5} 
changes sign between Model~2 (negative, without year FE) and Models~3--4 
(positive, with year FE), mirroring the pattern observed in the life 
satisfaction OLS and confirming that the income coefficient is sensitive 
to the inclusion of temporal controls.}

\subsection{Visual Representation of the Results}

As in the life satisfaction case, we illustrate how income, partisanship, and rurality jointly shape emotional well-being presenting, as an example, the model-predicted probabilities of happiness in 2022. 

Figure~\ref{fig:fig7} displays the model-predicted probability of expressing
above-{zero}  happiness ($P(\texttt{happiness}>0)$) across the Democratic–Republican
vote margin and three income strata (20th, 50th, and 80th percentiles of median
household income). On the other hand, Figure~\ref{fig:fig8} examines the role of income with
separate panels for Republican-leaning, competitive, and Democratic-leaning counties. 

Together, the two figures highlight that affective well-being is broadly 
positive, {negligibly related to partisanship, and shaped 
primarily by the structural urban--rural divide, consistent with the 
regression results in Table~\ref{tab:happiness}, where \texttt{margin} 
and \texttt{margin:rural} are not statistically significant.} In fact, 
across all panels of Figure~\ref{fig:fig7}, predicted probabilities are 
high (typically above 0.9), but a clear \emph{urban affective advantage} 
emerges{, which shows little systematic variation across 
the vote margin axis, confirming the negligible role of partisanship in 
shaping hedonic well-being.} The three income panels are nearly 
overlapping, confirming the negligible role of income. The same evidence 
is shown in Figure~\ref{fig:fig8}, where slopes along the income axis are 
nearly flat, indicating that expressed happiness varies little with 
household economic conditions. {The vote margin shows 
minimal moderating influence on the urban--rural gap, again consistent 
with the absence of a significant partisan effect in the regression 
models.}

{As in the life satisfaction figures, confidence intervals 
are omitted for visual clarity given the overlap across nine rurality 
codes; uncertainty estimates are reported in 
Table~\ref{tab:happiness_margins}.}

\begin{table}[!ht]
\centering
\caption{\label{tab:tab:ex_county_context}Socioeconomic context for extreme counties (ACS 2022 5-year). Rates shown as percentages; last row reports U.S. average.}
\centering
\begin{tabular}[t]{cllcrccccc}
\toprule
Year & State & County & Rural×Margin & Tweets & Poverty   & Unemp   & Black   & Native  & BA+  \\
  &   &   &   &   &   \% &   \% &   \% &   \% &   \%\\
\midrule
2014 & Alabama & Choctaw & 9.000 & 15168 & 19.1 & 5.2 & 40.0 & 0.2 & 13.0\\
2014 & Alabama & Perry & 8.000 & 22413 & 32.8 & 15.7 & 70.8 & 0.0 & 15.2\\
2014 & Alabama & Sumter & 8.000 & 38680 & 30.4 & 8.2 & 72.3 & 0.1 & 21.8\\
2014 & Alabama & Wilcox & 9.000 & 13189 & 26.8 & 10.7 & 70.2 & 0.1 & 11.6\\
2014 & Mississippi & Bolivar & 7.000 & 43579 & 31.8 & 7.4 & 63.2 & 0.0 & 27.2\\
\addlinespace
2014 & Mississippi & Carroll & 8.000 & 6366 & 21.8 & 6.0 & 35.8 & 0.0 & 18.0\\
2014 & Mississippi & Claiborne & 8.000 & 37489 & 30.4 & 9.0 & 85.8 & 0.0 & 22.3\\
2014 & Mississippi & Grenada & 7.000 & 34669 & 22.2 & 4.1 & 44.9 & 0.1 & 22.1\\
2014 & Mississippi & Humphreys & 8.000 & 6128 & 32.1 & 13.7 & 78.8 & 0.1 & 18.6\\
2014 & Mississippi & Issaquena & 9.000 & 272 & 20.6 & 12.1 & 70.4 & 0.0 & 5.2\\
\addlinespace
2014 & Mississippi & Jefferson & 9.000 & 21983 & 31.8 & 5.4 & 79.0 & 0.2 & 19.7\\
2014 & Mississippi & Leake & 8.000 & 10376 & 21.0 & 6.0 & 39.8 & 5.8 & 16.9\\
2014 & Mississippi & Montgomery & 9.000 & 11438 & 24.2 & 7.8 & 46.0 & 0.0 & 16.9\\
2014 & Mississippi & Quitman & 8.000 & 6187 & 30.2 & 12.2 & 71.9 & 0.3 & 14.4\\
2014 & Mississippi & Sharkey & 8.000 & 2038 & 26.9 & 12.8 & 72.5 & 0.0 & 16.0\\
\addlinespace
2014 & Mississippi & Sunflower & 7.000 & 12728 & 28.7 & 11.1 & 73.6 & 0.1 & 17.4\\
2014 & Mississippi & Tallahatchie & 9.000 & 15530 & 25.7 & 11.0 & 61.5 & 0.0 & 11.7\\
2014 & Mississippi & Yalobusha & 9.000 & 19740 & 20.5 & 5.0 & 38.4 & 0.2 & 12.3\\
2018 & New York & Allegany & 5.922 & 4781 & 17.1 & 7.3 & 1.7 & 0.1 & 24.5\\
2020 & South Dakota & Oglala Lakota & 7.121 & 2398 & 55.8 & 11.8 & 0.1 & 93.7 & 8.1\\
\addlinespace
2014 & Texas & McMullen & 8.000 & 3697 & 14.9 & 2.4 & 0.0 & 0.9 & 15.7\\
2018 & Washington & San Juan & 9.000 & 2920 & 11.0 & 3.2 & 0.6 & 0.6 & 52.6\\
2018 & Wisconsin & Lafayette & 7.996 & 244 & 10.5 & 2.4 & 0.5 & 0.2 & 19.9\\
\\
\hline
\\
\textbf{} & \textbf{U.S. average} & \textbf{} & \textbf{} & \textbf{} & \textbf{15.1} & \textbf{5.2} & \textbf{8.9} & \textbf{1.9} & \textbf{23.5}\\
\bottomrule
\end{tabular}
\end{table}

\section{Reconciling the Puzzle: Life Satisfaction \textit{vs.} 
Happiness}\label{sec:reconcile}

Table~\ref{tab:reconcile} compares the direction and relative strength 
of effects in the full weighted models for Twitter-based life satisfaction 
and happiness. Although the two indicators are positively correlated (see 
Table~\ref{tab:corr}), they respond in distinct ways to structural, 
temporal, and political factors. The contrast is especially clear in the 
spatial dimension.

\begin{table}[h!]
\centering
\caption{Comparative direction and strength of effects in the full models 
for Life Satisfaction and Happiness.}
\label{tab:reconcile}
\begin{tabular}{lcc}
\toprule
\textbf{Predictor} & \textbf{Life Satisfaction} & \textbf{Happiness} \\ 
\midrule
Rurality (\texttt{rural})          & $+++$ & $---$ \\
Partisan Margin (\texttt{margin})  & {$-$$^{\dagger}$} & {$0$} \\
Rural $\times$ Margin              & {$0$} & {$0$} \\
Household Income (\texttt{acs5})   & {$+$$^{\dagger}$} & $0$ \\
Year Effects (2016--2022)          & $+++$ & $---$ \\
\bottomrule
\end{tabular}
\begin{flushleft}
\footnotesize
\textit{Note:} Signs denote the direction of statistically significant 
effects in the full weighted models 
(Tables~\ref{tab:lifesat_margins} and~\ref{tab:happiness_margins}), 
based on logistic specifications with clustered standard errors (HC1) 
by county. The number of symbols reflects approximate magnitude: 
$+$ = weak, $++$ = moderate, $+++$ or $---$ = strong; $0$ = null 
effect. {$\dagger$ = effect present in the logistic 
specification but not robust across all specifications; see text for 
details.}
\end{flushleft}
\end{table}

\textbf{Rurality} exerts opposite effects on the two dimensions. Rural 
counties display systematically higher probabilities of \textit{life 
satisfaction} but lower probabilities of \textit{happiness}. This 
inversion suggests that rural residents express stronger long-term 
evaluative well-being, potentially reflecting stable life circumstances, 
lower expectations, and greater community cohesion, but show less 
momentary positive affect in their online communication. Urban counties, 
conversely, exhibit higher affective tone but lower evaluative 
contentment, consistent with the faster pace, denser social networks, 
and more expressive nature of metropolitan social media discourse. This 
urban--rural contrast underscores how the hedonic and cognitive components 
of well-being diverge sharply in their spatial geography.

{The \textbf{partisan vote margin} shows a suggestive 
negative association with \textit{life satisfaction} (AME $= -0.026$, 
$p = 0.009$) but no significant association with \textit{happiness} 
(AME $= -0.002$, $p = 0.831$). This asymmetry is theoretically 
informative: the partisan gradient in well-being, to the extent it 
exists, operates primarily through the evaluative dimension rather than 
hedonic affect. For \textit{life satisfaction}, the negative effect of 
\texttt{margin} is present but should be interpreted with caution, as 
it is sensitive to the clustering correction and reflects the 
near-time-invariant nature of county-level vote shares.
 The \texttt{rural}~$\times$~\texttt{margin} 
interaction is not statistically significant in either model under 
clustered standard errors, though the pattern of average marginal 
effects across rurality codes (Figure~\ref{fig:margin_rural}) suggests 
that the partisan gradient in life satisfaction attenuates at both ends 
of the urban--rural continuum. For happiness, the absence of both a 
main partisan effect and an interaction confirms that affective 
well-being is not structured along partisan lines once structural and 
temporal factors are accounted for.}

\textbf{Household income} contributes positively to life satisfaction 
in the logistic specification (AME $= 0.005$ per \$10{,}000, $p = 
0.004$) but shows no consistent relationship with happiness once 
evaluative well-being and other covariates are held constant. 
{However, this income effect is not robust across 
specifications: it is not statistically significant in the OLS model 
with year fixed effects (Table~\ref{tab:lifesatOLS}), suggesting that 
the income gradient in evaluative well-being is modest and 
specification-sensitive.} Evaluative well-being scales more strongly 
with economic prosperity, whereas hedonic well-being appears relatively 
income-insensitive once basic comfort levels are achieved. This 
asymmetry mirrors the ``satiation'' effect observed in classical 
well-being research, whereby positive affect plateaus beyond moderate 
income thresholds \citep{StevensonWolfers2008}.

Temporal dynamics further differentiate the two constructs. \textbf{Year 
effects} for happiness are large and negative in 2020--2022 (AMEs 
$\approx -0.020$ and $-0.041$), capturing the sharp deterioration in 
national mood during the COVID-19 pandemic and its aftermath. By 
contrast, life satisfaction exhibits {a somewhat positive}  
temporal variation (AMEs $\approx 0.080$ and $0.121$ for 2020 and 2022 
respectively), suggesting that it reflects a slower-moving cognitive 
evaluation of life circumstances less influenced by short-term shocks. 
This divergence reinforces the interpretation of happiness as a reactive, 
high-frequency indicator of social mood and life satisfaction as a more 
stable, structural indicator of enduring well-being.

Taken together, these findings demonstrate that our Twitter-based measures 
of \textit{happiness} and \textit{life satisfaction} capture complementary 
but distinct dimensions of subjective well-being. Life satisfaction aligns 
with the evaluative, cognitively integrated component and strongly 
influenced by rurality and {suggestively associated with} 
economic stability,  whereas happiness reflects the hedonic, affective 
component, which is more urban and responsive to 
national sentiment. Reconciling the two therefore requires acknowledging 
that geographic, {and} temporal contexts shape not only the 
level but also the \emph{type} of well-being expressed online, 
{while the role of partisan context, though present for 
evaluative well-being, is more modest and specification-sensitive than 
previously suggested.}

\subsection{Extreme Counties}

{Although the interaction between rurality and vote margin 
is not statistically significant under clustered standard errors in either 
model, the pattern of average marginal effects across rurality codes 
(Figure~\ref{fig:margin_rural}) suggests that the partisan gradient in 
life satisfaction varies with geographic context. In particular, the 
negative effect of \texttt{margin} on life satisfaction is somewhat 
stronger in semi-rural counties (codes 6--8) and attenuates at the 
extremes (codes 1 and 9). For happiness, no meaningful partisan gradient 
emerges across any rurality code. It is nonetheless informative to explore 
the characteristics of counties that sit at the intersection of high 
rurality and strong Democratic alignment, as these represent the contexts 
where the rural--partisan combination is most extreme and where the 
interpretation of our indicators requires the most care.}

In order to systematically identify these most extreme cases, we selected 
counties lying at the intersection of high rurality and strong Democratic 
alignment, defined as those with rurality scores above~7 (on the USDA 
rural--urban continuum) and electoral margins above~0.75. From this subset, 
we retained a single representative year per county corresponding to the 
maximum product of rurality and margin (see Section~\ref{sec:analysis}). 
To contextualize these counties socioeconomically, we merged their Federal 
Information Processing Standard (FIPS) codes with 2022 five-year American 
Community Survey (ACS) estimates from the U.S.~Census Bureau. For each 
county we extracted indicators of poverty, unemployment, racial composition, 
and educational attainment --- namely, the proportion of individuals below 
the poverty line (Table~B17001), the unemployment rate (Table~B23025), the 
shares of Black or African American and Native American residents 
(Table~B02001), and the percentage of adults with a bachelor's degree or 
higher (Table~B15003). These variables summarize the structural conditions 
associated with the most rural and Democratic-leaning contexts 
(Table~\ref{tab:tab:ex_county_context}).

The most extreme counties in our data cluster overwhelmingly in the Deep 
South, particularly across the Mississippi Delta and Alabama's Black Belt, 
alongside a few rural outliers in New York, Wisconsin, Washington, and 
South Dakota (Table~\ref{tab:tab:ex_county_context}). The southern cluster 
displays a distinctive pattern of persistent socioeconomic disadvantage: 
poverty rates between 25\% and~33\%, unemployment roughly twice the 
national mean (5.2\%), and a predominantly Black population exceeding 
60--80\% in several counties. Educational attainment is markedly lower 
than the U.S.~average (23.5\% with a bachelor's degree or higher), with 
most counties below~20\%. Historically, the Black Belt has exhibited 
entrenched racial and spatial inequalities since the post-Reconstruction 
era, coupled with low levels of civic participation and institutional trust 
\citep{duBois1899philadelphia,key1949southern,partridge2007persistent}.

In this context, {the well-being signals observed in our 
models for these counties likely reflect a deeper social cleavage rather 
than straightforward evaluative or hedonic states.} Rather than capturing 
subjective well-being per se, social-media expressions of ``happiness'' 
and ``life satisfaction'' in these counties may represent in-group identity 
and affective polarization within culturally homogeneous and historically 
segregated environments. The inclusion of Oglala~Lakota County, one of the 
poorest and most marginalized Native American communities in the United 
States (poverty rate~56\%, 94\% Native population, and only~8\% college 
educated), further underscores how economic deprivation and historical 
exclusion can shape online affective signals independently of conventional 
well-being measures.

While most extreme counties share patterns of deprivation, a few outliers 
appear at the opposite end of the socioeconomic spectrum. McMullen County 
(Texas) and San~Juan County (Washington) record poverty and unemployment 
far below the national averages (15.1\% and~5.2\%, respectively), with 
San~Juan displaying the highest educational attainment in the sample 
(52.6\% with a bachelor's degree or more). Lafayette County (Wisconsin) 
is similarly prosperous and predominantly White. These counties illustrate 
that strong rurality and political homogeneity can coexist with widely 
different material conditions, from entrenched poverty and racial 
segregation in the Deep South to relative affluence and educational 
advantage in isolated northern or coastal enclaves. The reversal in the 
life-satisfaction gradient thus reflects not a single socioeconomic pattern 
but a shared feature of \emph{social insularity}, whether shaped by 
exclusion or by privilege.

\section{Caveats and Robustness of the Findings}\label{sec:risks}

Our approach leverages social-media text classified by a fine-tuned 
language model and aggregated to the county level. This design provides 
exceptional temporal resolution and spatial coverage, but it also entails 
potential inferential risks. Below we discuss the main caveats and 
summarize the robustness checks implemented to mitigate them.

\paragraph{Coverage and selection bias.}
Twitter/X users are not representative of county populations, 
over-representing younger, urban, and politically engaged groups, and 
tweet density varies widely across space 
\citep{pew2019twitterusers,pew2019politics,mislove2011understanding,
hecht2014tale,hawelka2014geo}. {Importantly, our goal is 
not to estimate population-level prevalence of well-being, but to 
document patterns in expressed self-perception across geographic groups. 
A differential propensity to tweet need not mechanically invalidate 
comparisons in expressed content, although who speaks and how they speak 
remains a genuine limitation that cannot be fully resolved with 
observational data. Overturning the main findings would require assuming 
that the direction of any selection bias in self-expression differs 
systematically across groups in a way that is correlated with our 
outcomes. Results should therefore  be interpreted as patterns in the 
population of geolocated tweets, not as a direct representation of the 
residential population.}
\emph{Mitigations:} (i) precision weighting via inverse standard errors 
to down-weight noisy counties, with weights strongly correlated with 
county population on a log scale ($r = 0.70$ for life satisfaction, 
$r = 0.75$ for happiness); (ii) modeling at the county level with year 
fixed effects to isolate contextual variation; (iii) confirmation that 
temporal and structural patterns (rural advantage in life satisfaction, 
urban advantage in happiness, and pandemic-era declines) align with 
external benchmarks.

\paragraph{Geolocation and mapping error.}
Imperfect or shifting geotags can misassign tweets across counties.
\emph{Mitigations:} use of the Harvard CGA geocoding pipeline with 
rigorous county mapping.

\paragraph{LLM classification error and drift.}
General-purpose models may misinterpret sarcasm, idioms, or local 
cultural context, and class priors may drift over time.
\emph{Mitigations:} (i) fine-tuning on manually annotated exemplars 
aligned with the intended constructs; (ii) human-in-the-loop validation; 
(iii) freezing model weights for each time slice to prevent drift; 
(iv) revalidation across random tweet subsamples and alternative prompts 
\citep{finetuning2024}.

\paragraph{Construct validity.}
Text-based happiness and life satisfaction must correspond to the hedonic 
and evaluative dimensions they are intended to capture.
\emph{Evidence:} (i) convergent validity: moderate correlations between 
the two indicators, consistent with theory; (ii) discriminant validity: 
opposite signs for the rural--urban gradient (rural advantage in life 
satisfaction, urban advantage in happiness); (iii) external validity: 
expected associations with income, partisanship, and the pandemic period.

\paragraph{Model specification and ecological inference.}
Dichotomizing continuous scores can obscure variation, and aggregation 
may introduce ecological bias.
\emph{Mitigations:} (i) {the threshold at zero is a 
substantively meaningful boundary between negative and positive expressed 
sentiment, grounded in the measurement design rather than an arbitrary 
statistical split;} (ii) replication using OLS models with clustered 
standard errors yields consistent coefficients and signs, confirming that 
logistic results are not artifacts of dichotomization; (iii) robustness 
to alternative specifications including continuous outcomes and interaction 
terms; (iv) controls for year effects.

\paragraph{Statistical inference.}
{The panel structure of the data,  with the same counties 
observed across multiple years, induces within-county serial correlation 
that can lead to underestimated standard errors. We address this by 
reporting heteroskedasticity-robust standard errors clustered at the 
county level (HC1) for all specifications 
\citep{wooldridge2010econometric,ArelBundock2024}. Under this correction, the core 
results, i.e., the rural advantage in life satisfaction, the urban advantage in 
happiness, and the pandemic-era year effects, remain strongly significant. 
The partisan margin (\texttt{margin}) and its interaction with rurality 
(\texttt{margin:rural}) lose statistical significance in the OLS 
specifications, consistent with the near-time-invariant nature of 
county-level vote shares. We therefore treat partisan effects as 
suggestive rather than conclusive.}

\paragraph{Platform and policy shocks.}
Algorithmic changes or moderation practices may distort sentiment trends.
\emph{Mitigations:} inclusion of year fixed effects to absorb 
national-level platform shocks.

\paragraph{Generalizability.}
Findings pertain to affective and evaluative expressions on Twitter/X, 
not to the full U.S.\ population.
\emph{Mitigations:} results are explicitly framed as complementary to 
survey-based measures.

\paragraph{Ethical and privacy considerations.}
Although data are aggregated, ethical use remains essential.
\emph{Practices:} (i) county-level aggregation; (ii) no user-level 
identifiers.

\bigskip

{Overall, the consistency of effect directions across 
logistic and OLS specifications with clustered standard errors, the 
theoretical plausibility of the rural--urban inversion, and the robustness 
of temporal patterns (year fixed effects for happiness and life 
satisfaction) all suggest that these indicators capture genuine dimensions 
of collective well-being rather than artifacts of platform behavior or 
modeling choices. The more modest evidence on partisan gradients, while 
suggestive, should be interpreted with appropriate caution given its 
sensitivity to model specification.}

 \section{Conclusions}\label{sec:conclusions}

Aesop's ancient fable of the town mouse and the country mouse 
\citep{aesop2002} is quite explicit in portraying urban life as offering 
greater material opportunities, yet marked by heightened precariousness 
and frenetic pace, whereas rural life is depicted as poorer but more 
serene. This depiction aligns well with the differing trends in 
\emph{happiness} and \emph{life satisfaction} discussed in the preceding 
pages. More specifically, our research helps clarify why prior U.S. studies \citep{atherton2024,winters2017,glaeser2006} reached mixed findings: rural counties 
appear more content in long-term evaluative terms, while urban counties 
exhibit higher short-term affective tone. The distinction aligns with the 
classical hedonic--evaluative dichotomy in well-being theory.

The analysis leverages 2.6~billion geo-referenced social-media posts and 
a fine-tuned generative AI classifier to produce county-year indicators of 
life satisfaction and happiness for the United States, 2014--2022. By 
combining language-model--based inference with statistical modeling, we 
identify robust and theory-consistent patterns in digital expressions of 
subjective well-being. Three main conclusions stand out: \textit{i)} rural 
and urban well-being diverge by dimension, \textit{ii)} {
political context shows a suggestive but specification-sensitive 
association with evaluative well-being}, and \textit{iii)} temporal shocks 
dominate happiness but not life satisfaction.

{After controlling for temporal shocks and applying 
clustered standard errors, Democratic-leaning counties show a suggestive 
negative association with life satisfaction (AME $= -0.026$, $p = 0.009$) 
but no significant association with happiness (AME $= -0.002$, $p = 
0.831$). The interaction between rurality and partisan margin is not 
statistically significant under clustered standard errors in either model, 
though the pattern of average marginal effects across rurality codes 
(Figure~\ref{fig:margin_rural}) suggests that the partisan gradient in 
life satisfaction attenuates at both ends of the urban--rural continuum. 
What might appear as a simple ``blue versus red'' difference in well-being 
is therefore more nuanced: partisan effects on evaluative well-being are 
present but modest, specific to the evaluative dimension, and sensitive to 
model specification. No meaningful partisan gradient emerges for hedonic 
affect.}

Year fixed effects for 2020--2022 are sharply negative for happiness, 
coinciding with the pandemic and its aftermath, whereas life satisfaction 
shows more muted and positive movements. This temporal dissociation is 
precisely what we would expect if the two indicators capture distinct 
affective layers: happiness as a volatile, mood-sensitive construct, and 
life satisfaction as a slower-moving evaluative judgment.

\medskip
Methodologically, we have shown that \emph{social-media--based well-being 
measures can make scientific sense}. They behave as theory anticipates 
(distinct yet correlated dimensions), respond plausibly to macro shocks, 
and reveal interpretable cross-sectional structure (rural--urban contrasts 
and temporal dynamics). {The robustness of the core 
findings, i.e.,  the rural advantage in life satisfaction, the urban advantage 
in happiness, and the pandemic-era declines, holds across logistic and 
OLS specifications with clustered standard errors, supporting the 
validity of these indicators as measures of collective expressed 
well-being.} Compared to traditional surveys, they offer three unique 
advantages: (i)~\emph{granularity}: coverage of every U.S.\ county; 
(ii)~\emph{continuity}: annual measurement across nine years; and 
(iii)~\emph{responsiveness}: ability to detect fast-moving societal 
shocks.

For researchers, these indicators enable high-resolution monitoring of 
subjective well-being and the study of how local context shapes both 
affective and evaluative dimensions of flourishing. {The 
distinction between evaluative and hedonic well-being proves consequential: 
partisan and income gradients, while suggestive for life satisfaction, do 
not extend to happiness, underscoring the importance of measuring and 
analyzing the two dimensions separately rather than aggregating them into 
a single well-being index.} For policymakers, they suggest differentiated 
strategies: improving \emph{evaluative} well-being in urban areas may 
require addressing cost-of-living and stability concerns, while fostering 
\emph{affective} well-being in rural areas may call for strengthening 
social connection, access to services, and opportunity, especially for 
low-income communities.
\section*{Funding Declaration}
This work received in-kind financial support in the form of GPU computing resources provided by the Advanced Cyberinfrastructure Coordination Ecosystem: Services \& Support (ACCESS) program, under allocations ``SOC250043: The Geography of Human Flourishing” and ``SOC250007: The Geography of Hate and Human Flourishing: Exploring Global Attitudes Towards Immigration using Geotweets.”

\section*{Data Availability}
Replication data and scripts used to perform the analysis in this study can be found in the Harvard Dataverse repository at this link \url{https://doi.org/10.7910/DVN/BVCEER}.

\clearpage

\eject

\appendix 

\section*{Appendix A. Additional Tables and Figures.}

\begin{table}[!htbp] \centering 
  \caption{Logit estimates of $\Pr(\texttt{lifesat} > 0)$ by county 
(2014--2022). All models are weighted by the inverse standard deviation of 
life satisfaction estimates. Covariates are added sequentially: rurality, 
partisan margin, income (\texttt{acs5} in \$10,000), year fixed effects, 
and their interaction.}
  \label{tab:lifesat} 
\begin{tabular}{@{\extracolsep{5pt}}lcccc} 
\\[-1.8ex]\hline 
\hline \\[-1.8ex] 
 & \multicolumn{4}{c}{\textit{Dependent variable:}} \\ 
\cline{2-5} 
\\[-1.8ex] & \multicolumn{4}{c}{$\Pr(\texttt{lifesat} > 0)$} \\ 
\\[-1.8ex] & (1) & (2) & (3) & (4)\\ 
\hline \\[-1.8ex] 
 rural & 0.219$^{***}$ & 0.333$^{***}$ & 0.312$^{***}$ & 0.291$^{***}$ \\ 
  & (0.021) & (0.024) & (0.030) & (0.037) \\ 
  & & & & \\ 
 acs5 ($\times$ \$10,000) &  & 0.328$^{***}$ & 0.123$^{**}$ & 0.120$^{**}$ \\ 
  &  & (0.037) & (0.042) & (0.042) \\ 
  & & & & \\ 
 year = 2016 &  &  & 4.559$^{***}$ & 4.581$^{***}$ \\ 
  &  &  & (1.348) & (1.342) \\ 
  & & & & \\ 
 year = 2018 &  &  & 0.669$^{*}$ & 0.696$^{*}$ \\ 
  &  &  & (0.290) & (0.299) \\ 
  & & & & \\ 
 year = 2020 &  &  & 1.912$^{***}$ & 1.952$^{***}$ \\ 
  &  &  & (0.218) & (0.233) \\ 
  & & & & \\ 
 year = 2022 &  &  & 2.719$^{***}$ & 2.754$^{***}$ \\ 
  &  &  & (0.239) & (0.258) \\ 
  & & & & \\ 
 margin & $-$0.233 & $-$0.395$^{**}$ & $-$0.457$^{**}$ & $-$0.309 \\ 
  & (0.124) & (0.128) & (0.161) & (0.239) \\ 
  & & & & \\ 
 happiness & 34.852$^{***}$ & 35.499$^{***}$ & 39.006$^{***}$ & 39.050$^{***}$ \\ 
  & (0.851) & (0.880) & (1.603) & (1.602) \\ 
  & & & & \\ 
 margin $\times$ rural &  &  &  & $-$0.052 \\ 
  &  &  &  & (0.063) \\ 
  & & & & \\ 
 Constant & $-$9.628$^{***}$ & $-$12.195$^{***}$ & $-$12.982$^{***}$ & $-$12.955$^{***}$ \\ 
  & (0.212) & (0.352) & (0.407) & (0.407) \\ 
  & & & & \\ 
\hline \\[-1.8ex] 
Observations & 13,642 & 13,640 & 13,640 & 13,640 \\ 
Log Likelihood & $-$24,697.730 & $-$23,821.440 & $-$20,999.530 & $-$20,992.990 \\ 
Akaike Inf. Crit. & 49,403.450 & 47,652.880 & 42,017.060 & 42,005.970 \\ 
\hline 
\hline \\[-1.8ex] 
\textit{Note:}  & \multicolumn{4}{r}{Clustered standard errors (HC1) by county in parentheses. $^{*}$p$<$0.05; $^{**}$p$<$0.01; $^{***}$p$<$0.001.} \\ 
\end{tabular} 
\end{table} 

\begin{table}[!htbp]
\centering
\caption{Marginal effect of partisan vote margin 
(\texttt{margin}) on $\Pr(\texttt{lifesat} > 0)$ by rurality code, 
from the full model (Model~4). Clustered standard errors (HC1) by county 
via \texttt{marginaleffects} \citep{ArelBundock2024}.} 
\label{tab:margin_by_rural}
\begin{tabular}{rrrrlrr}
  \toprule
Rural.code & AME & SE & z & p & lower & upper \\ 
  \midrule
1 & -0.0150 & 0.0081 & -1.84 & 0.066 & -0.0309 & 0.0010 \\ 
  2 & -0.0180 & 0.0074 & -2.44 & 0.015 & -0.0324 & -0.0036 \\ 
  3 & -0.0204 & 0.0071 & -2.85 & 0.004 & -0.0344 & -0.0064 \\ 
  4 & -0.0268 & 0.0092 & -2.91 & 0.004 & -0.0448 & -0.0088 \\ 
  5 & -0.0272 & 0.0102 & -2.67 & 0.008 & -0.0472 & -0.0072 \\ 
  6 & -0.0321 & 0.0133 & -2.41 & 0.016 & -0.0581 & -0.0060 \\ 
  7 & -0.0361 & 0.0167 & -2.16 & 0.031 & -0.0688 & -0.0033 \\ 
  8 & -0.0323 & 0.0162 & -2.00 & 0.046 & -0.0640 & -0.0006 \\ 
  9 & -0.0308 & 0.0169 & -1.83 & 0.067 & -0.0639 & 0.0022 \\ 
   \bottomrule
\end{tabular}
\end{table}

\begin{table}[!htbp] \centering 
  \caption{Weighted logistic regression of above-{zero} happiness ($P(\texttt{happiness} > 0)$) on rurality, partisan vote margin, household income, and year fixed effects. }
  \label{tab:happiness} 
\begin{tabular}{@{\extracolsep{5pt}}lcccc} 
\\[-1.8ex]\hline 
\hline \\[-1.8ex] 
 & \multicolumn{4}{c}{\textit{Dependent variable:}} \\ 
\cline{2-5} 
\\[-1.8ex] & \multicolumn{4}{c}{$\Pr(\texttt{happiness} > 0)$} \\ 
\\[-1.8ex] & (1) & (2) & (3) & (4)\\ 
\hline \\[-1.8ex] 
 rural & $-$0.420$^{***}$ & $-$0.584$^{***}$ & $-$0.542$^{***}$ & $-$0.500$^{***}$ \\ 
  & (0.114) & (0.102) & (0.089) & (0.121) \\ 
  & & & & \\ 
 acs5 ($\times$ \$10,000) &  & $-$0.402$^{***}$ & $-$0.069 & $-$0.005 \\ 
  &  & (0.042) & (0.058) & (0.065) \\ 
  & & & & \\ 
 year = 2016 &  &  & $-$5.310$^{***}$ & $-$5.269$^{***}$ \\ 
  &  &  & (0.688) & (0.687) \\ 
  & & & & \\ 
 year = 2018 &  &  & $-$2.208$^{***}$ & $-$2.085$^{***}$ \\ 
  &  &  & (0.533) & (0.480) \\ 
  & & & & \\ 
 year = 2020 &  &  & $-$3.549$^{***}$ & $-$3.462$^{***}$ \\ 
  &  &  & (0.436) & (0.478) \\ 
  & & & & \\ 
 year = 2022 &  &  & $-$4.515$^{***}$ & $-$4.443$^{***}$ \\ 
  &  &  & (0.945) & (0.795) \\ 
  & & & & \\ 
 margin & $-$0.943 & $-$0.944 & $-$1.031 & $-$2.644 \\ 
  & (0.623) & (0.609) & (1.058) & (2.456) \\ 
  & & & & \\ 
 lifesat & 10.962$^{***}$ & 12.273$^{***}$ & 13.608$^{**}$ & 13.432$^{**}$ \\ 
  & (2.964) & (3.155) & (4.733) & (4.533) \\ 
  & & & & \\ 
 margin $\times$ rural &  &  &  & 0.318 \\ 
  &  &  &  & (0.346) \\ 
  & & & & \\ 
 Constant & 8.406$^{***}$ & 11.577$^{***}$ & 12.407$^{***}$ & 11.971$^{***}$ \\ 
  & (0.626) & (0.688) & (0.698) & (0.630) \\ 
  & & & & \\ 
\hline \\[-1.8ex] 
Observations & 13,642 & 13,640 & 13,640 & 13,640 \\ 
Log Likelihood & $-$15,442.910 & $-$14,778.870 & $-$12,586.150 & $-$12,422.410 \\ 
Akaike Inf. Crit. & 30,893.820 & 29,567.740 & 25,190.310 & 24,864.820 \\ 
\hline 
\hline \\[-1.8ex] 
\textit{Note:}  & \multicolumn{4}{r}{Clustered standard errors (HC1) by county in parentheses. $^{*}$p$<$0.05; $^{**}$p$<$0.01; $^{***}$p$<$0.001.} \\ 
\end{tabular} 
\end{table}

\begin{figure}
    \centering
    \includegraphics[width=0.75\linewidth]{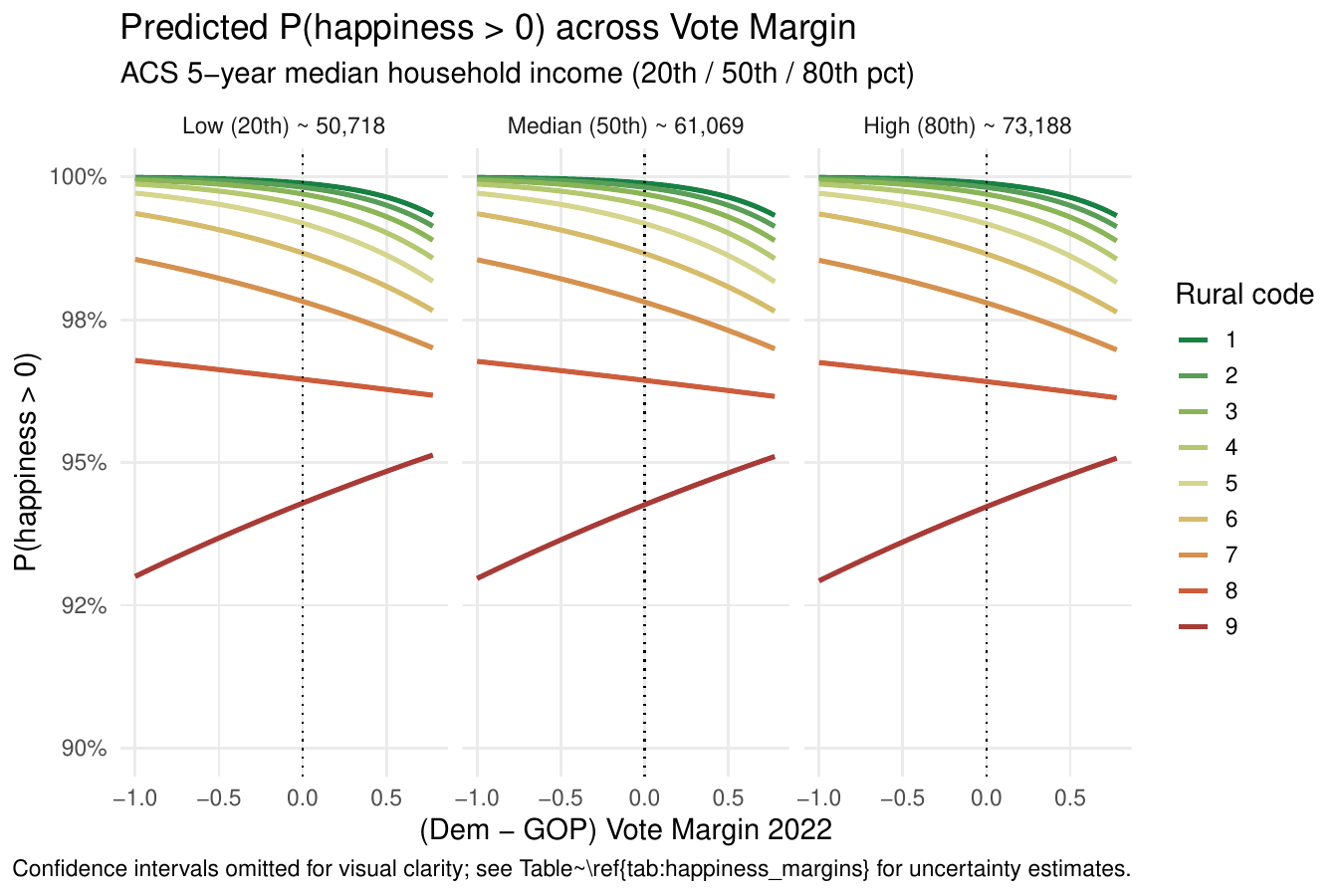}
\caption{Predicted probability of expressing positive happiness 
($P(\texttt{happiness}>0)$) across the Democratic--Republican vote margin 
for U.S.\ counties in 2022, based on the full model in 
equation~\eqref{eq:happiness_full}. Panels correspond to the 20th, 50th, 
and 80th percentiles of median household income (ACS 5-year estimates). 
Lines represent rurality codes (1--9). Happiness levels are uniformly high 
but decline slightly with rurality, {consistently with the (although non statistically significant) \texttt{margin} 
coefficient in Table~\ref{tab:happiness}. Confidence intervals are omitted 
for visual clarity given the overlap across nine rurality codes; uncertainty 
estimates are reported in Table~\ref{tab:happiness_margins}.}}
    \label{fig:fig7}
\end{figure}

\begin{figure}
    \centering
    \includegraphics[width=0.75\linewidth]{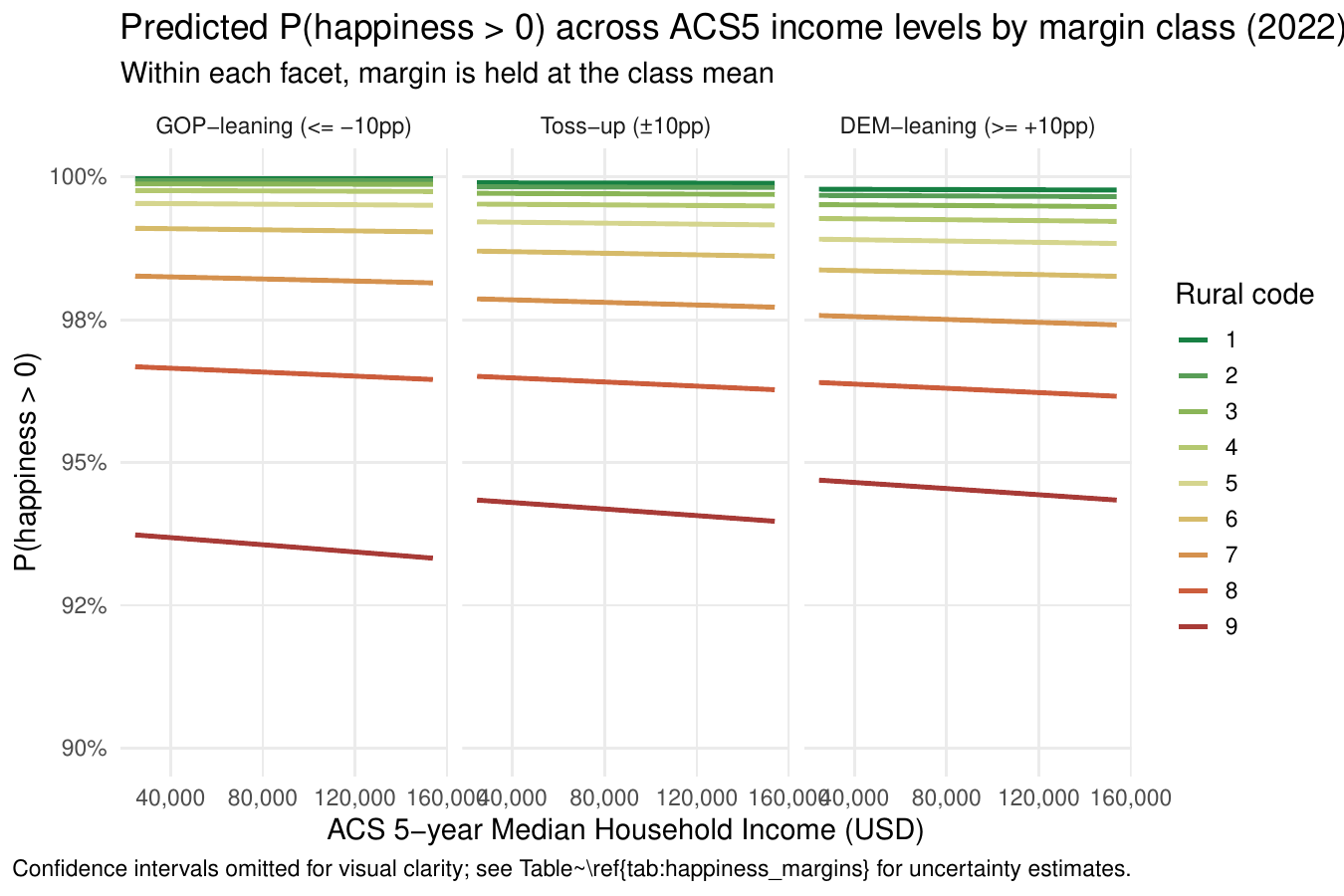}
\caption{Predicted probability of expressing positive happiness 
($P(\texttt{happiness}>0)$) across median household income levels for 
U.S.\ counties in 2022, based on the full model in 
equation~\eqref{eq:happiness_full}. Panels correspond to 
Republican-leaning ($\leq -10$ pp), toss-up ($\pm 10$ pp), and 
Democratic-leaning ($\geq +10$ pp) counties. Within each panel the 
partisan margin is held at the class mean. Lines represent rurality 
codes (1--9). Happiness increases modestly with income but remains 
consistently higher in urban counties across all partisan contexts. 
{Political orientation has minimal influence once income 
and rurality are controlled, consistent with the non-significant 
\texttt{margin} coefficient in Table~\ref{tab:happiness} and 
underscoring the predominance of the urban affective advantage over 
partisan context in shaping hedonic well-being. Confidence intervals 
are omitted for visual clarity given the overlap across nine rurality 
codes; uncertainty estimates are reported in 
Table~\ref{tab:happiness_margins}.}}
    \label{fig:fig8}
\end{figure}

\begin{table}[!htbp] \centering 
  \caption{OLS estimates of \texttt{lifesat} by county (2014--2022). 
All models are weighted by the inverse standard deviation of life satisfaction 
estimates. Covariates are added sequentially: rurality, partisan margin, 
income (\texttt{acs5} in \$10,000), year fixed effects, and their 
interaction. Clustered standard errors (HC1) by county in parentheses.} 
  \label{tab:lifesatOLS} 
\begin{tabular}{@{\extracolsep{5pt}}lcccc} 
\\[-1.8ex]\hline 
\hline \\[-1.8ex] 
 & \multicolumn{4}{c}{\textit{Dependent variable:}} \\ 
\cline{2-5} 
\\[-1.8ex] & \multicolumn{4}{c}{\texttt{lifesat}} \\ 
\\[-1.8ex] & (1) & (2) & (3) & (4)\\ 
\hline \\[-1.8ex] 
 rural & 0.004$^{***}$ & 0.005$^{***}$ & 0.005$^{***}$ & 0.004$^{***}$ \\ 
  & (0.0003) & (0.0003) & (0.0003) & (0.0004) \\ 
  & & & & \\ 
 acs5 ($\times$ \$10,000) &  & 0.004$^{***}$ & 0.0003 & 0.0003 \\ 
  &  & (0.0003) & (0.0003) & (0.0003) \\ 
  & & & & \\ 
 year = 2016 &  &  & 0.176$^{***}$ & 0.176$^{***}$ \\ 
  &  &  & (0.011) & (0.011) \\ 
  & & & & \\ 
 year = 2018 &  &  & 0.004 & 0.004 \\ 
  &  &  & (0.003) & (0.003) \\ 
  & & & & \\ 
 year = 2020 &  &  & 0.023$^{***}$ & 0.023$^{***}$ \\ 
  &  &  & (0.001) & (0.001) \\ 
  & & & & \\ 
 year = 2022 &  &  & 0.052$^{***}$ & 0.052$^{***}$ \\ 
  &  &  & (0.002) & (0.002) \\ 
  & & & & \\ 
 margin & 0.007$^{***}$ & 0.006$^{***}$ & $-$0.003$^{**}$ & $-$0.002 \\ 
  & (0.001) & (0.001) & (0.001) & (0.002) \\ 
  & & & & \\ 
 happiness & 1.207$^{***}$ & 1.205$^{***}$ & 0.970$^{***}$ & 0.970$^{***}$ \\ 
  & (0.005) & (0.005) & (0.019) & (0.019) \\ 
  & & & & \\ 
 margin $\times$ rural &  &  &  & $-$0.001 \\ 
  &  &  &  & (0.0005) \\ 
  & & & & \\ 
 Constant & $-$0.274$^{***}$ & $-$0.299$^{***}$ & $-$0.267$^{***}$ & $-$0.266$^{***}$ \\ 
  & (0.001) & (0.002) & (0.003) & (0.003) \\ 
  & & & & \\ 
\hline \\[-1.8ex] 
Observations & 13,642 & 13,640 & 13,640 & 13,640 \\ 
R$^{2}$ & 0.925 & 0.926 & 0.943 & 0.943 \\ 
Adjusted R$^{2}$ & 0.925 & 0.926 & 0.943 & 0.943 \\ 
Residual Std. Error & 0.233 (df = 13638) & 0.231 (df = 13635) & 0.202 (df = 13631) & 0.202 (df = 13630) \\ 
\hline 
\hline \\[-1.8ex] 
\textit{Note:}  & \multicolumn{4}{r}{$^{*}$p$<$0.05; $^{**}$p$<$0.01; $^{***}$p$<$0.001.} \\ 
\end{tabular} 
\end{table}

\begin{table}[!htbp] \centering 
   \caption{Linear regression of $\texttt{happiness}$ on rurality, partisan vote margin, household income, and year fixed effects. Results are in line with Table~\ref{tab:happiness}.}
  \label{tab:happinessOLS} 
\begin{tabular}{@{\extracolsep{5pt}}lcccc} 
\\[-1.8ex]\hline 
\hline \\[-1.8ex] 
 & \multicolumn{4}{c}{\textit{Dependent variable:}} \\ 
\cline{2-5} 
\\[-1.8ex] & \multicolumn{4}{c}{\texttt{happiness}} \\ 
\\[-1.8ex] & (1) & (2) & (3) & (4)\\ 
\hline \\[-1.8ex] 
 rural & $-$0.002$^{***}$ & $-$0.003$^{***}$ & $-$0.002$^{***}$ & $-$0.002$^{***}$ \\ 
  & (0.0002) & (0.0002) & (0.0002) & (0.0002) \\ 
  & & & & \\ 
 acs5 ($\times$ \$10,000) &  & $-$0.002$^{***}$ & 0.001$^{***}$ & 0.001$^{***}$ \\ 
  &  & (0.0003) & (0.0002) & (0.0002) \\ 
  & & & & \\ 
 year = 2016 &  &  & 0.095$^{***}$ & 0.095$^{***}$ \\ 
  &  &  & (0.006) & (0.006) \\ 
  & & & & \\ 
 year = 2018 &  &  & 0.049$^{***}$ & 0.049$^{***}$ \\ 
  &  &  & (0.003) & (0.003) \\ 
  & & & & \\ 
 year = 2020 &  &  & $-$0.022$^{***}$ & $-$0.022$^{***}$ \\ 
  &  &  & (0.001) & (0.001) \\ 
  & & & & \\ 
 year = 2022 &  &  & $-$0.018$^{***}$ & $-$0.018$^{***}$ \\ 
  &  &  & (0.001) & (0.001) \\ 
  & & & & \\ 
 margin & $-$0.004$^{***}$ & $-$0.003$^{***}$ & $-$0.001 & $-$0.003 \\ 
  & (0.001) & (0.001) & (0.001) & (0.001) \\ 
  & & & & \\ 
 lifesat & 0.797$^{***}$ & 0.797$^{***}$ & 0.661$^{***}$ & 0.660$^{***}$ \\ 
  & (0.002) & (0.002) & (0.008) & (0.008) \\ 
  & & & & \\ 
 margin $\times$ rural &  &  &  & 0.001 \\ 
  &  &  &  & (0.0004) \\ 
  & & & & \\ 
 Constant & 0.222$^{***}$ & 0.238$^{***}$ & 0.201$^{***}$ & 0.200$^{***}$ \\ 
  & (0.001) & (0.002) & (0.002) & (0.002) \\ 
  & & & & \\ 
\hline \\[-1.8ex] 
Observations & 13,642 & 13,640 & 13,640 & 13,640 \\ 
R$^{2}$ & 0.962 & 0.963 & 0.971 & 0.971 \\ 
Adjusted R$^{2}$ & 0.962 & 0.963 & 0.971 & 0.971 \\ 
Residual Std. Error & 0.246 (df = 13638) & 0.244 (df = 13635) & 0.214 (df = 13631) & 0.214 (df = 13630) \\ 
\hline 
\hline \\[-1.8ex] 
\textit{Note:}  & \multicolumn{4}{r}{Clustered standard errors (HC1) by county in parentheses. $^{*}$p$<$0.05; $^{**}$p$<$0.01; $^{***}$p$<$0.001.} \\ 
\end{tabular} 
\end{table}

\clearpage

\eject

\section*{Appendix B. Construction, Fine-Tuning, and Validation of 
Tweet-Based Well-Being Indicators}\label{sec:appendix}

This appendix summarizes how the tweet-level indicators of \emph{life 
satisfaction} and \emph{happiness} were derived using fine-tuned large 
language models, and further aggregated to the county level. A full 
methodological account of the broader Human Flourishing Geographic Index 
(HFGI) dataset, including additional well-being dimensions, validation 
benchmarks, and reproducibility resources --- is provided in 
\cite{iacus2025}. Here we describe only the two components analyzed in 
this article: the evaluative (\texttt{lifesat}) and hedonic 
(\texttt{happiness}) dimensions.

\subsection*{B.1 Fine-tuning and classification pipeline}

We trained and validated a fine-tuned large language model (LLM) to 
classify short social-media texts according to multiple flourishing-related 
constructs. The training corpus comprised 4,581 manually annotated tweets, 
selected from an initial pool of 10,000 messages sampled across topics, 
years, and geographies to ensure linguistic diversity. Each tweet was 
independently labeled by human coders following standardized 
dimension-specific guidelines. For both \texttt{lifesat} and 
\texttt{happiness}, coders indicated whether the tweet referred to the 
construct and, if so, whether the expressed tone was \textit{low}, 
\textit{medium}, or \textit{high}. These labels were used to fine-tune an 
instruction-following generative model (LLaMA family), optimizing 
cross-entropy loss across the three classes.

The fine-tuned model produced a JSON-formatted classification for each 
tweet, returning both the target label and confidence score. Evaluation on 
a held-out test set yielded accuracies around 0.87 depending on the 
dimension, with consistent precision--recall balance. Post-hoc validation 
involved manual review of ambiguous cases to verify conceptual alignment 
with the human flourishing framework (see \cite{iacus2025} for details).

{To clarify how the two constructs are operationalized in 
practice, we provide a brief description of the linguistic content 
associated with each classification level. Tweets classified as expressing 
\emph{life satisfaction} (\texttt{lifesat}) at a \textit{high} level 
typically contain evaluative, reflective language about one's life as a 
whole. For example, expressions of contentment with one's living 
situation, community, or life trajectory, or forward-looking statements 
of stability and fulfillment. A \textit{low} classification corresponds 
to the opposite: dissatisfaction with one's life circumstances in an 
overall, evaluative sense. A \textit{medium} classification captures 
ambivalent or moderately positive evaluative language. Tweets classified 
as expressing \emph{happiness} (\texttt{happiness}) at a \textit{high} 
level reflect immediate positive affect or joy in response to a specific 
moment or event, whereas \textit{low} corresponds to sadness or emotional 
distress in the short term.}

{The evaluative--hedonic distinction produces interpretable 
cross-combinations. A tweet expressing stable contentment with one's life 
overall (high \texttt{lifesat}) while conveying a subdued or melancholic 
mood in the moment (low \texttt{happiness}) exemplifies the dissociation 
that motivates our analytical framework. Conversely, a tweet expressing 
joy at an immediate event (high \texttt{happiness}) need not imply 
overall life contentment (high \texttt{lifesat}).} 
{To further illustrate these distinctions, we provide 
below two synthetic examples, constructed to be representative of the 
classification scheme without reproducing actual user content.
\begin{itemize}
\item \textit{High \texttt{lifesat}, low \texttt{happiness}}: 
``Grateful for everything I have built over the years. 
Still, today feels heavy and I can't quite shake this sadness.''
This tweet expresses an overall evaluative contentment with one's life 
trajectory (\texttt{lifesat} = high) while conveying a negative 
immediate affective state (\texttt{happiness} = low).
\item \textit{Low \texttt{lifesat}, high \texttt{happiness}}: 
``Not where I want to be in life, but tonight with friends laughing 
until we cried -- couldn't ask for more right now.''
This tweet expresses dissatisfaction with one's life circumstances 
overall (\texttt{lifesat} = low) alongside a strong positive hedonic 
experience in the moment (\texttt{happiness} = high).
\end{itemize}
These examples illustrate the evaluative--hedonic dissociation that 
motivates our analytical framework. The full codebook with synthetic 
examples for all 48 dimensions is available in \citet{iacus2025}, 
Section~8.2.}

\subsection*{B.2 Scaling and county-level aggregation}

We later transformed the three-level scale (``low'', ``medium'', 
``high'') as: $-1$ (\textit{negative}), $+0.5$ (\textit{somewhat 
positive}), $+1$ (\textit{positive}) and set $0$ if a tweet does not 
contain expressions related to life satisfaction or happiness 
(\textit{not-present}). Notice that low maps to the opposite concept. 
For example ``happiness'' = ``low'' means ``sadness''. {
The threshold at zero therefore corresponds to the natural boundary 
between negative and positive expressed sentiment --- a substantively 
meaningful dichotomization grounded in the measurement design, not an 
arbitrary statistical split.}

Each classified tweet was geolocated to a U.S.\ county using the Harvard 
CGA pipeline. Tweets with ambiguous or low-confidence geotags were 
excluded. For every county--month--dimension triple, we computed the mean 
and standard deviation per geographical region (county/state) and temporal 
frequency (month/year). For descriptive purposes, we also computed the 
total tweet volume and the share of tweets classified in each dimension 
(``salience''), which are provided in the accompanying data release.

\subsection*{B.3 Validation and interpretation}

Validation followed the multi-stage procedure described in 
\cite{iacus2025}. \textit{Convergent validity} was confirmed by moderate 
positive correlations between \texttt{lifesat} and \texttt{happiness} 
across counties. \textit{Discriminant validity} was evidenced by their 
opposite spatial gradients (rural advantage in life satisfaction versus 
urban advantage in happiness) consistent with the evaluative--hedonic 
distinction in well-being theory. \textit{External validity} was assessed 
by comparing other indicators in the broader HFGI dataset with independent 
benchmarks. In particular, \citet{iacus2025} validate the 
\texttt{depression} indicator against the CDC's Behavioral Risk Factor 
Surveillance System (BRFSS) county-level estimates of frequent mental 
distress, finding positive and significant associations once rural--urban 
differences in expression propensity are accounted for. The 
\texttt{lifesat} and \texttt{happiness} indicators analyzed here are 
constructed using the same pipeline and validated through convergent and 
discriminant validity checks as described above.

These two dimensions represent a subset of the larger HFGI corpus, which 
also encompasses civic, moral, social, and health-related domains of 
flourishing. For complete documentation of the fine-tuning data, 
validation benchmarks, and aggregation code, readers are referred to 
\cite{iacus2025}.

\clearpage
\eject

\bibliography{sample}

\end{document}